\documentclass[twocolumn]{aastex631}

\usepackage{savesym}
\savesymbol{tablenum}
\usepackage{siunitx}
\restoresymbol{SIX}{tablenum}

\usepackage[version=4]{mhchem}
\usepackage{graphicx,epsf}
\usepackage{subfigure}
\usepackage{graphicx}	
\usepackage{amsmath}	
\usepackage{amssymb}	
\usepackage{newtxtext,newtxmath}
\usepackage{siunitx}
\usepackage{footnote}

\newcommand{\Msun}{{\rm M}_\odot}
\newcommand{\Rsun}{{\rm R}_\odot}

\newcommand{\kms}{\textrm{km}\,\textrm{s}^{-1}}

\def\sn{{SN~2024ggi}}
\newcommand{\mdot}{M$_{\odot}$~yr$^{-1}$}

\DeclareRobustCommand{\ion}[2]{\relax\ifmmode\ifx\testbx\f@series{\mathbf{#1\,\mathsc{#2}}}\else{\mathrm{#1\,\mathsc{#2}}}\fi\else\textup{#1\,{\mdseries\textsc{#2}}}\fi}

\newcommand{\code}[1]{\texttt{#1}}

\def\cmfgen{{\code{CMFGEN}}}
\def\sumo{{\code{SUMO}}}
\def\extrass{{\code{EXTRASS}}}
\def\sedona{{\code{SEDONA}}}

\usepackage[scale=0.94]{tgbonum}
\usepackage[mode=text]{siunitx}

\makeatletter
\DeclareTextCompositeCommand{\r}{OT1}{A}{%
  \leavevmode\vbox{%
    \offinterlineskip
    \ialign{\hfil##\hfil\cr\char23\cr\noalign{\kern-1.15ex}A\cr}%
  }%
}
\makeatother

\shorttitle{Supernova Infrared Emission Lines}
\shortauthors{Jacobson-Gal\'an et al.}

\begin{document}

\title{Mapping 3-D Explosive Nucleosynthesis in Type II Supernova 2024ggi with Infrared Emission Lines}

\correspondingauthor{Wynn Jacobson-Gal\'{a}n (he, him, his)}
\email{wynnjg@caltech.edu}

\author[0000-0002-3934-2644]{W.~V.~Jacobson-Gal\'{a}n}
\altaffiliation{NASA Hubble Fellow}
\affiliation{Cahill Center for Astrophysics, California Institute of Technology, MC 249-17, 1216 E California Boulevard, Pasadena, CA, 91125, USA}

\author[0000-0003-0599-8407]{L.~Dessart}
\affil{Institut d’Astrophysique de Paris, CNRS-Sorbonne Université, 98 bis boulevard Arago, F-75014 Paris, France}
\affil{French-Chilean Laboratory for Astronomy, IRL 3386, CNRS and Instituto de Astrofísica, Pontificia Universidad Católica de Chile, Casilla 306, Santiago, Chile.}

\author[0000-0003-1938-9282]{D.~Vartanyan}
\affil{Department of Physics, University of Idaho, ID 83843, USA}

\begin{abstract}

We present analysis and modeling of optical and infrared (IR) spectroscopy of the Type II supernova (SN~II) 2024ggi obtained with ground-based instruments and the {\it James Webb Space Telescope (JWST)} at phases of $\sim265$--400~days. The near- and mid-IR spectra reveal diverse iron-group emission-line morphologies, including double-peaked profiles in [\ion{Ni}{i}]~3.119 and 11.998~$\mu$m, [\ion{Fe}{ii}]~1.644 and 17.931~$\mu$m, and [\ion{Co}{i}]~12.255~$\mu$m, alongside Gaussian profiles in [\ion{Ni}{ii}]~1.939~$\mu$m, [\ion{Co}{ii}]~10.520~$\mu$m, and [\ion{Ni}{i}]~7.505 and 11.304~$\mu$m. These differences imply both chemical inhomogeneity and aspherical ionization of inner ejecta, consistent with expectations from the $^{56}$Ni bubble effect. Modeling of double-peaked profiles supports an ejecta distribution with polar enhancements as large as $\sim7$ for Ni/Co/Fe-rich material and $\sim2$ for intermediate-mass elements. LTE estimates imply a stable Ni mass of $M_{\rm Ni}\approx2.1\times10^{-3}~\Msun$, but electron densities near critical values indicate departures from LTE. Comparisons to non-LTE radiative transfer models favor a progenitor mass of $\sim12$--15.2~$\Msun$. We show that a simple mapping between elemental mass distribution and projected velocity reproduces line profiles produced in a \cmfgen\ radiative transfer calculation. We apply this property to 3-D neutrino-driven explosion simulations and predict Ni emission profiles for varying viewing angles. We find that only energetic 3-D explosion models of high-mass progenitors reproduce the observed extent of Ni mixing in SN~2024ggi, conflicting with progenitor masses inferred from radiative transfer models. These results demonstrate the utility of resolved nebular IR lines as direct probes of the 3-D distribution of explosively synthesized material in core-collapse SNe.

\end{abstract}


\section{Introduction} \label{sec:intro}

Constraining the distribution of metals within the ejecta of core-collapse supernovae (SNe) is a central goal of SN theory as it encodes key diagnostics of the neutrino-driven explosion mechanism (e.g., \citealt{Janka12, Burrows21}). Among the synthesized elements, radioactive $^{56}$Ni is particularly important. It is produced during explosive silicon burning following shock revival, and its spatial distribution and asymptotic velocity structure retain an imprint of the explosion dynamics, including the degree of asymmetry and the evolution of forward and reverse shocks (e.g., see \citealt{Kifonidis00, kifonidis_2003, wongwathanarat_15_3d, gabler_3dsn_21}). In addition to $^{56}$Ni, core-collapse SNe synthesize stable neutron-rich isotopes such as $^{58}$Ni and stable cobalt isotopes (e.g., $^{59}$Co), while iron is predominantly present as $^{56}$Fe, both primordial and as the decay product of $^{56}$Co \citep{seitenzahl14}. Importantly, significant amounts of these stable isotopes are produced in the same regions as $^{56}$Ni, making their emission lines valuable tracers of the original nickel distribution.

The distribution of nickel-rich material has been a key diagnostic of neutrino-driven explosions since the type II supernova (SN II) 1987A. It probes both large- and small-scale asymmetries and reflects the underlying hydrodynamics of the explosion (e.g., \citealt{wongwathanarat_15_3d, gabler_3dsn_21, Vartanyan25, Giudici25}). The extent of mixing depends sensitively on explosion energy, the amplitude of seed perturbations, explosion asymmetry, and the progenitor’s internal structure. For example, more energetic and asymmetric explosions tend to produce stronger outward mixing of heavy elements \citep{Vartanyan25, Vartanyan25SBO, Giudici25}. In addition, hydrodynamic instabilities—particularly Rayleigh–Taylor instabilities triggered as the shock propagates through composition interfaces—further amplify mixing and asymmetry \citep{mueller_1991, kifonidis_2003, kifonidis_2006, janka_2016}. The so-called $^{56}$Ni bubble effect, operating on timescales of weeks, also contributes to the redistribution of material \citep{Li93, Woosley05, gabler_3dsn_21}. These processes are now routinely captured in modern three-dimensional neutrino-driven explosion simulations, which follow the evolution from core collapse through shock breakout and into the homologous expansion phase \citep{gabler_3dsn_21}. In parallel, increasingly sophisticated radiative-transfer codes (e.g., \cmfgen, \extrass, \sedona, \sumo) can post-process such simulations and produce synthetic spectra for direct comparison with observations \citep{sedona,Jerkstrand12, Jerkstrand14, Dessart21, vanbaal_extrass_23, Barmentloo24, Dessart25IR, vanBaal25}, achieving promising agreement with recent datasets (e.g., SN~2024ggi; \citealt{Dessart24ggi}).

Observationally, however, constraints on the three-dimensional chemical structure of SN ejecta have been indirect. Evidence for enhanced $^{56}$Ni  mixing in SN~1987A was inferred from the early high-energy emission and rapid rise of the optical light curve (e.g., \citealt{matz_1988, shigeyama_1988, chupp_1989, arnett_1989, mccray_1993}) as well as the ``Bochum event'' \citep{Hanuschik88, Utrobin95, Wang02}. Later, radio observations enabled three-dimensional mapping of C-, O-, and Si-rich material through molecular emission \citep{abellan_87A_17, Wesson26}. Similar reconstructions were achieved in supernova remnants such as Cassiopeia A \citep{Grefenstette14, Milisavljevic15, Milisavljevic24} and SNR 0540$-$69.3 \citep{Larsson26}, although in these cases the structure has been altered by interaction of the reverse shock with the ejecta. Direct spectroscopic probes of young SN ejecta have remained limited, particularly in the infrared (IR), where existing late-time ($>$1-2~years post-explosion) observations have generally been of low spectral resolution (e.g., SN~1987A; \cite{wooden_87A_ir_93}; SN~2004et; \cite{kotak_04et_09}). To date, only recently have a small number of Type II SNe been observed at moderate resolution with the \textit{James Webb Space Telescope} (\textit{JWST}), including SNe~1987A \citep{Jones23, Larsson23, Kavanagh26}, 2022acko \citep{Shahbandeh24}, 2023ixf \citep{DerKacy25, Medler25, wjg25d} and 2024ggi \citep{Baron25, Dessart24ggi, Mera26}.

The advent of {\it JWST} enables a more direct approach to probing the three-dimensional distribution of SN ejecta provided the use of sufficient spectral resolution. Infrared emission lines of metals suffer significantly less from line blending than optical transitions and are more optically-thin, particularly at wavelengths beyond $\sim 2\,\mu\mathrm{m}$, allowing clearer identification and interpretation of individual features. These include transitions from \ion{Ni}{i/ii}, \ion{Co}{i/ii}, \ion{Fe}{i/ii}, \ion{Ne}{ii}, and \ion{Ar}{ii}, which provide complementary diagnostics to optical lines and offer direct insight into the distribution of iron-peak and intermediate-mass elements. In addition, IR spectroscopy enables the study of molecular emission, particularly CO, whose fundamental band near $\sim 5\,\mu\mathrm{m}$ provides a powerful tracer of the C/O-rich zones and requires the wavelength coverage and sensitivity of \textit{JWST} \citep{Park25, Dessart24ggi, Medler25, Mera26}. At late times, IR emission also becomes essential for quantifying the energy budget of the system as ejecta–circumstellar interaction increasingly dominates the luminosity \citep{Dessart23b,wjg25d}. Recent \textit{JWST} observations of Type Ia supernovae have already demonstrated the potential of IR spectroscopy for constraining explosion mechanisms and progenitor systems \citep{derkacy_21aefx_23,kwok_21aefx_23,kwok_22pul_24, Blondin23}. Extending these capabilities to core-collapse SNe offers a unique opportunity to directly probe the geometry, composition, and physical conditions of the inner ejecta, and to connect observations with predictions of modern multi-dimensional explosion models.

In this paper, we analyze and model nebular-phase optical and IR spectra of the SN~II 2024ggi, which showed multi-wavelength observational evidence for ejecta interaction with confined circumstellar material (CSM) in its first $\sim$week \citep{wjg24b, Shrestha24ggi, Chen24, Zhang24ggi, Pessi24, Yang25}. Following the fading of its early-time ``IIn-like'' emission lines, \sn{} evolved as a standard SN~II-P/L with a defined light curve ``plateau'' and radioactive decay powered ``tail.'' Late-time optical and ultraviolet spectroscopy \citep{Bostroem25, Hueichap26} and X-ray observations \citep{Ferdinand26} confirmed ongoing SN ejecta interaction with more distant CSM with a ``wind-like'' density profile that is consistent with a progenitor mass loss rate of $\dot M \approx 6\times 10^{-5}$~\mdot ($v_w = 20~\kms$).

\begin{figure*}
\centering
\includegraphics[width=\textwidth]{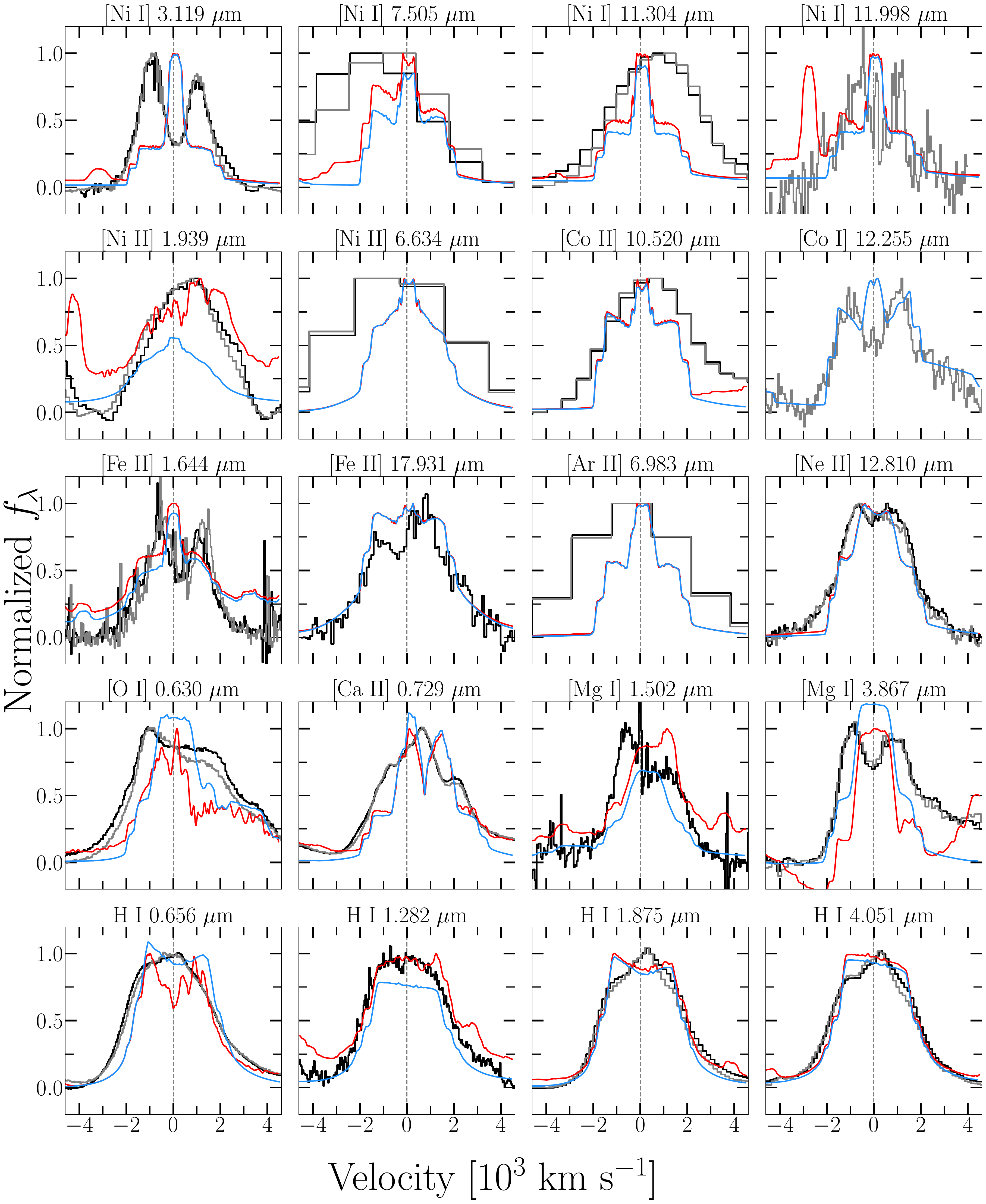}
\caption{Nebular optical/IR emission line velocities of SN~2024ggi at $\delta t = 265 - 291$~days (black) and $\delta t = 386 - 417$~days (gray), compared to s15p2 {\tt CMFGEN} model (red; \citealt{Dessart24ggi}). Shown in blue are the individual element contributions within the s15p2 model. The mid-IR coverage with {\it JWST} enables quantification of emission line profile shapes across varying ionization states and different iron-group and intermediate-mass elements.  \label{fig:all} }
\end{figure*}

In this work, we use a redshift-independent host-galaxy distance of $7.24 \pm 0.20$~Mpc \citep{Saha06} and redshift of $z = 0.002215$ \citep{wjg24b}. All phases reported in this paper are with respect to time of first light ($\delta t$) MJD $60410.80 \pm 0.34$~days. All data are corrected for MW and host-galaxy reddening values of \textit{E(B-V)} = 0.07~mag and $E(B-V)_{\textrm{host}} = 0.084 \pm 0.018$~mag. In \S\ref{sec:obs} we describe the data used, and in \S\ref{sec:analysis} we present analysis and model comparisons of \sn{}'s spectroscopic properties. Finally, in \S\ref{sec:discussion} we present a physical picture of the inner ejecta and explosive nucleosynthesis as well as discuss the current limitations and future avenues for such studies. Conclusions are drawn in \S\ref{sec:conclusion}. All uncertainties are quoted at the 68\% confidence level (c.l.) unless otherwise stated.

\section{Observations} \label{sec:obs}

For this study, we analyze all \sn{} spectra presented in \cite{Dessart24ggi}, which covers optical/IR wavelengths from $\sim 0.31 - 21~\mu$m.  This dataset includes Keck/LRIS optical ($0.31-1.028~\mu$m) spectra at $\delta t = 291 \ \& \ 417$~days and Keck/NIRES near-IR ($0.96 - 2.47~\mu$m) spectra at $\delta t = 277 \ \& \ 400$~days. {\it JWST} NIRSpec observations at $\delta t = 286.3 \ \& \ 386.5$~days covers a wavelength range of $1.66 - 5.27~\mu$m with a resolving power of $R \approx 1000$. {\it JWST} MIRI low-resolution (LRS) observations at $\delta t = 265.1 \ \& \ 389.5$~days covers a wavelength range of $\sim 5 - 14~\mu$m with a resolving power of $R \approx 40-160$. {\it JWST} medium resolution (MRS) MIRI spectra which were obtained at $\delta t = 265.3 \ \& \ 389.5$~days with the SHORT grating setting for all 4 channels cover wavelength ranges of $4.90- 5.74$, $7.51-8.77$, $11.55-13.47$, and $17.70-20.95$ with resolving power of $R \approx 3500 - 1700$.  

\section{Analysis}\label{sec:analysis}

\subsection{SN~2024ggi Optical-Infrared Nebular Emission}\label{subsec:spec}

\begin{figure*}
\centering
\subfigure{\includegraphics[width=0.32\textwidth]{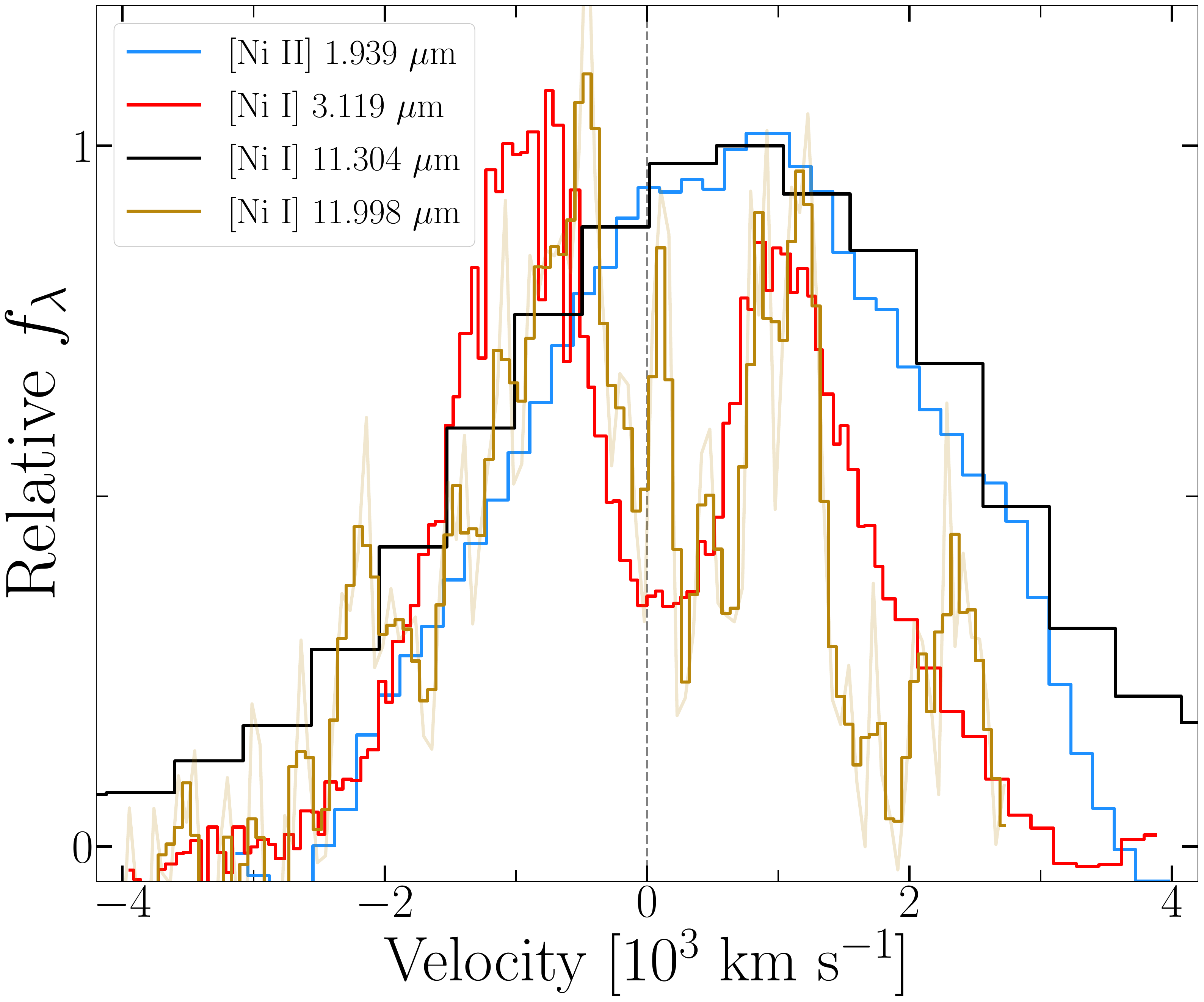}}
\subfigure{\includegraphics[width=0.32\textwidth]{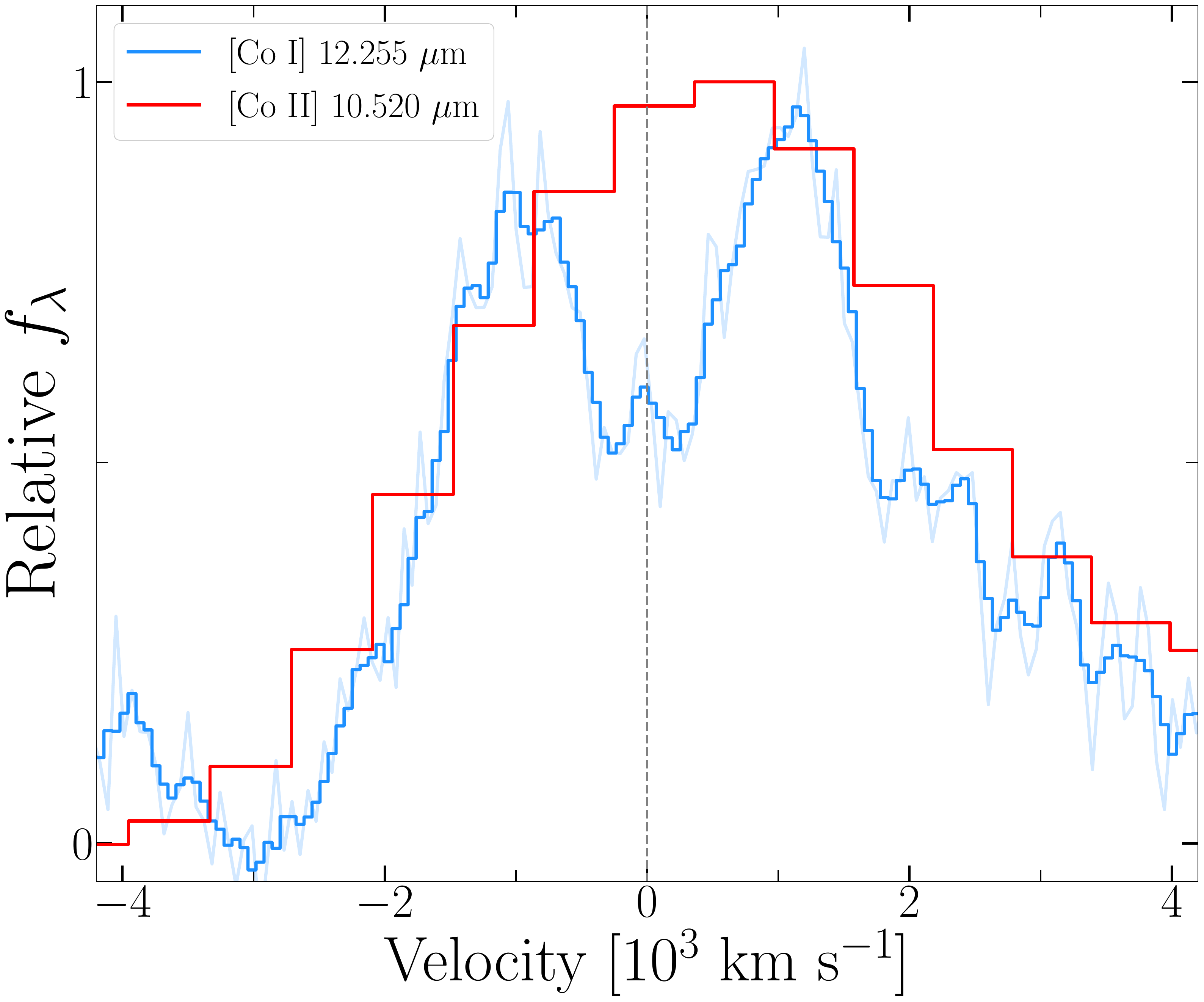}}
\subfigure{\includegraphics[width=0.32\textwidth]{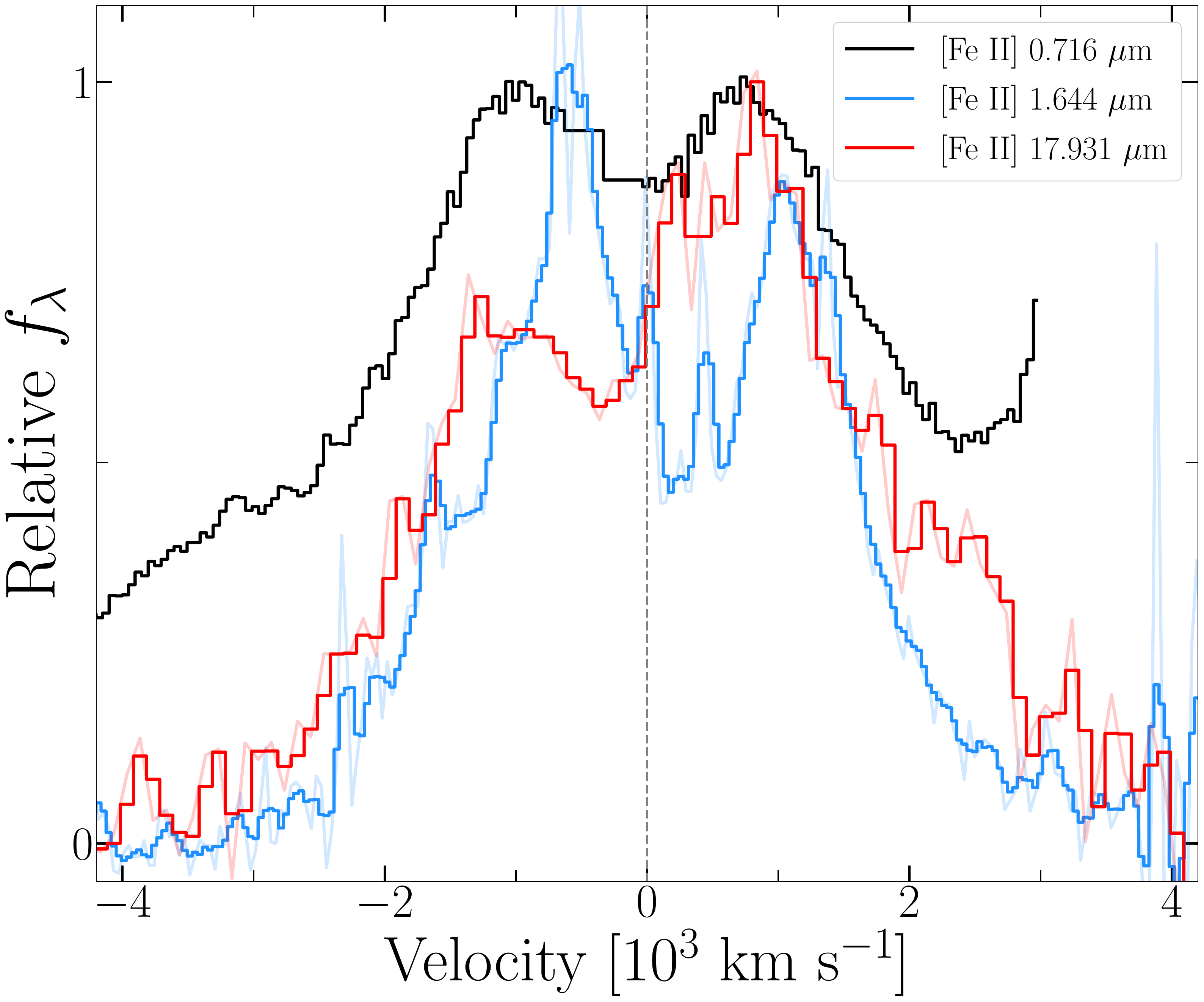}}
\caption{SN~2024ggi nebular infrared emission line velocities of Nickel ({\it Left}), Cobalt ({\it Middle}), and Iron ({\it Right}) ions. A prominent double-peaked line profile morphology is observed in \ion{Ni}{i},\ion{Co}{i} and \ion{Fe}{ii} transitions. However, some neutral and ionized Ni and Co transitions exhibit a Gaussian-like profile, peaked at red-ward velocities. \label{fig:IGEvels} }
\end{figure*}

\subsubsection{Line Profile Morphology}

In Figure \ref{fig:all}, we show iron group element (IGE), intermediate mass element (IME) and hydrogen emission line velocities in SN~2024ggi optical, NIR and MIR spectra at $\delta t \approx 265.3 - 400$~days. We also plot the full s15p2 model spectrum at $\delta t = 400$~days from \cite{Dessart24ggi} as well as line profiles for individual elements in order to quantify line overlap. The s15p2 model has $R_{\star} = 860~\Rsun$, $E_k = 0.84$~B, $M_{\rm ej} = 10.95~\Msun$, $M({\rm ^{56}Ni)} = 0.063~\Msun$ and maximum $^{56}$Ni velocity of $1997~\kms$. At these late-time phases, [\ion{Ni}{i/ii}] emission arises from stable $^{58}$Ni that, in the s15p2 model, is predicted to be co-located with $^{56}$Ni and its decay products (e.g., $^{56}$Co, $^{56}$Fe), all of which were created during explosive nucleosynthesis after core bounce and shock revival \citep{Woosley02, Sukhbold16}. As shown in Figures \ref{fig:all} and \ref{fig:IGEvels}, Ni emission lines show diverse structure across different ionization states. For example, [\ion{Ni}{i}] $\lambda$3.119~$\mu$m shows clear double-peaked emission with deep central indentation and blue-/red-ward peaks at $\sim 1000~\kms$. A similar emission line morphology is observed in [\ion{Ni}{i}] $\lambda$11.998~$\mu$m, but is not reflected in the [\ion{Ni}{i}] $\lambda$11.304~$\mu$m and [\ion{Ni}{ii}] $\lambda$1.939~$\mu$m, both of which show a Gaussian-like structure with redshifted emission peaking at $\sim 1000~\kms$. However, as demonstrated by the s15p2 model decomposition, there is line overlap from Ca and H transitions that contaminate the red-side of [\ion{Ni}{ii}] $\lambda$1.939~$\mu$m. We note that [\ion{Ni}{i}] $\lambda$7.505~$\mu$m and [\ion{Ni}{ii}] $\lambda$6.634~$\mu$m lines also show a single Gaussian shape but given the poor resolution of the MIRI LRS spectrum, the true shape of these lines is unfortunately not constrained. In contrast to the diversity observed across Ni lines, the profiles in s15p2 model are all approximately the same shape --- the narrow central peak is largely an artifact of the shuffled-shell structure in which the innermost ejecta layers is by design Ni-rich \citep{Dessart21, Dessart25IR}.

\begin{figure*}
\centering
\subfigure{\includegraphics[width=0.33\textwidth]{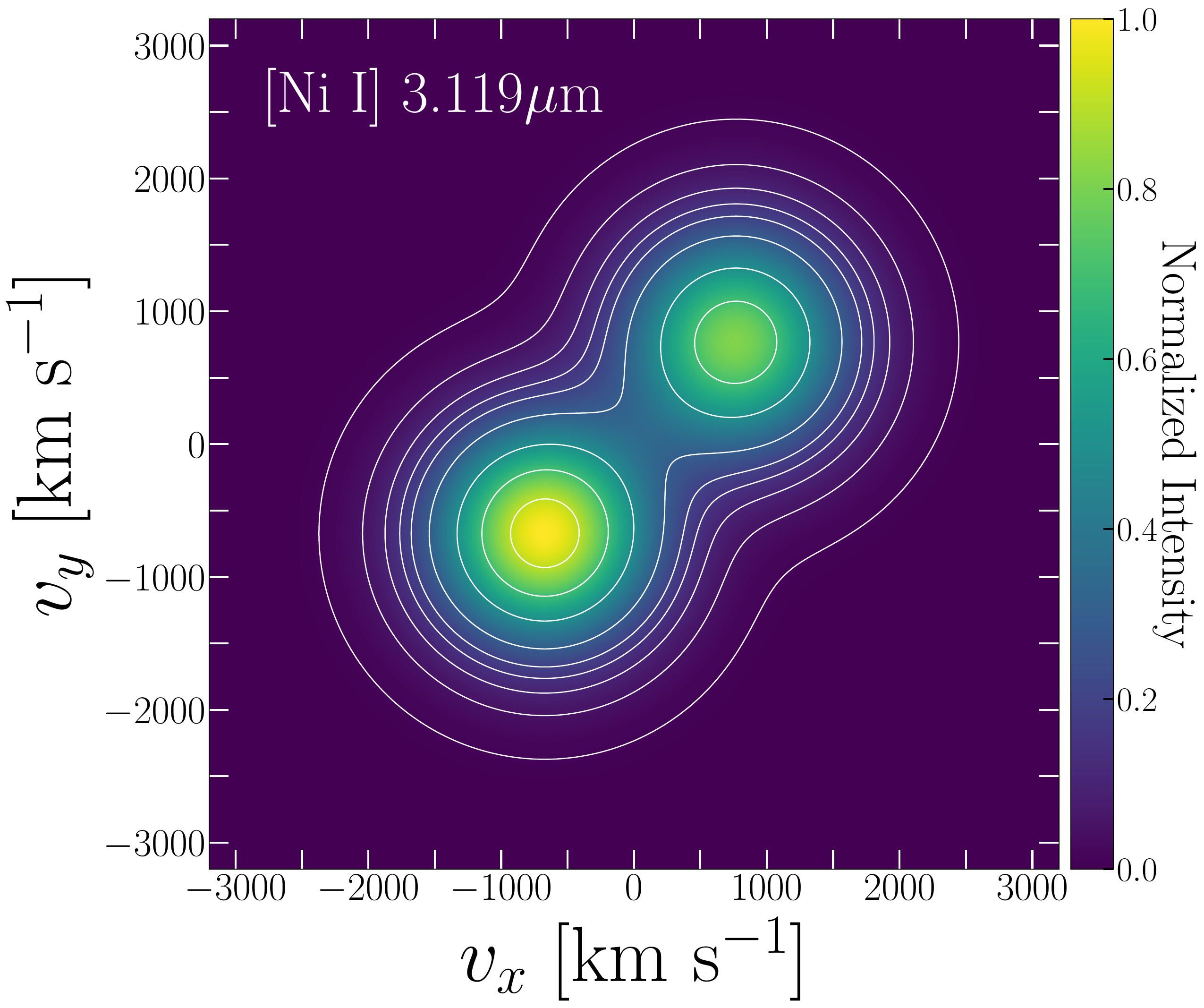}}
\subfigure{\includegraphics[width=0.33\textwidth]{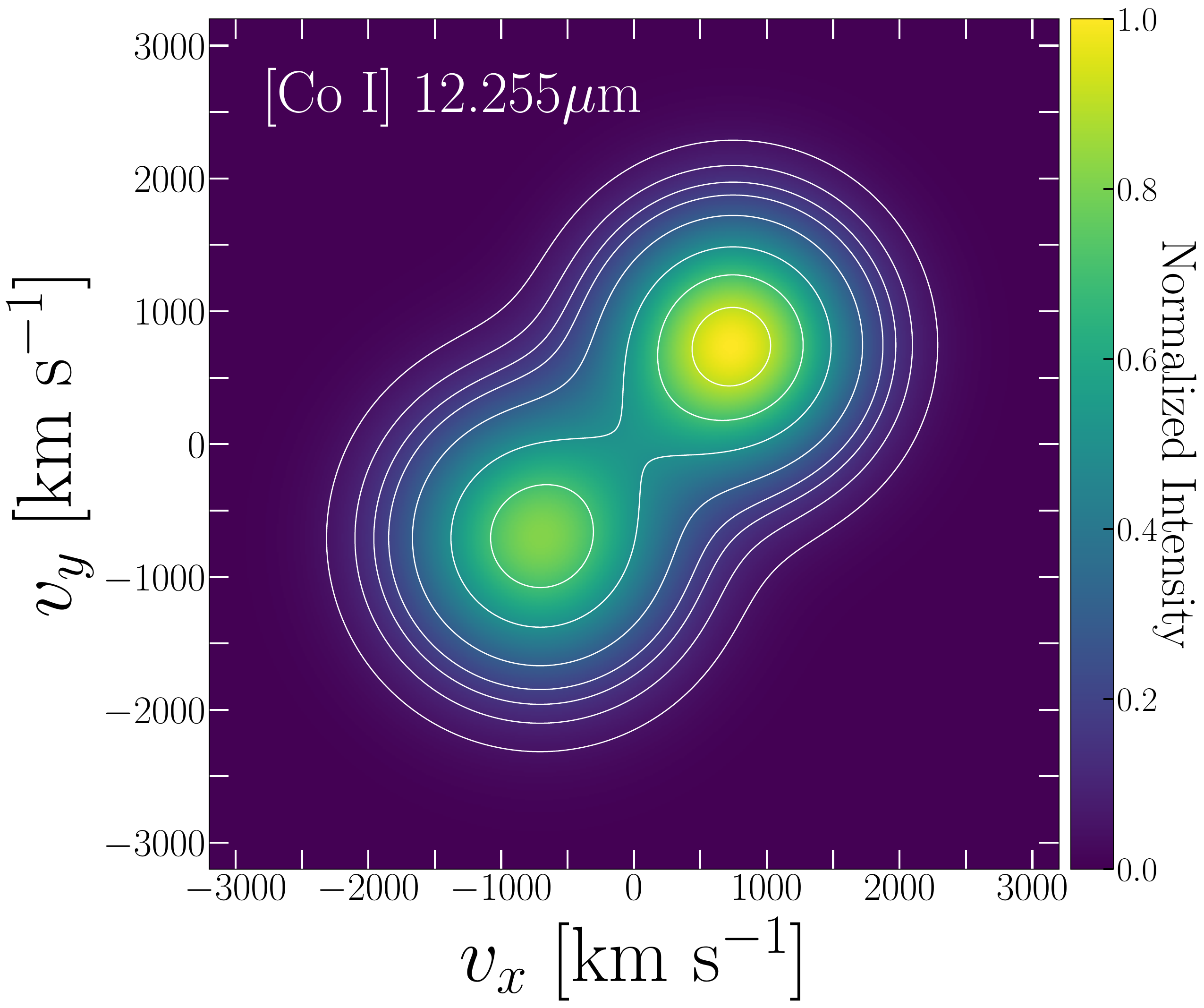}}
\subfigure{\includegraphics[width=0.33\textwidth]{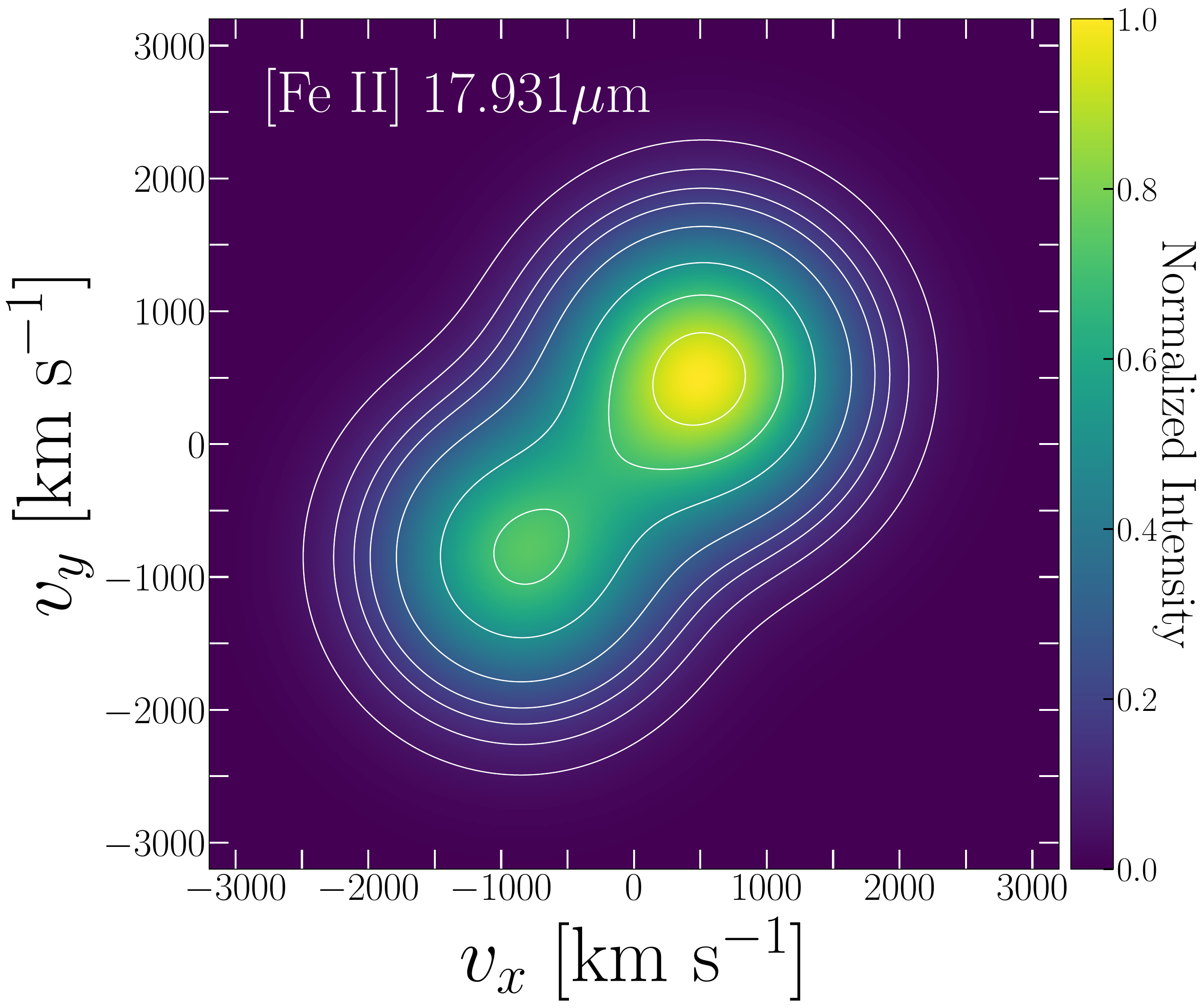}}\\
\noindent
\makebox[0.33\textwidth][c]{\hspace*{-2.8mm}\includegraphics[width=0.27\textwidth]{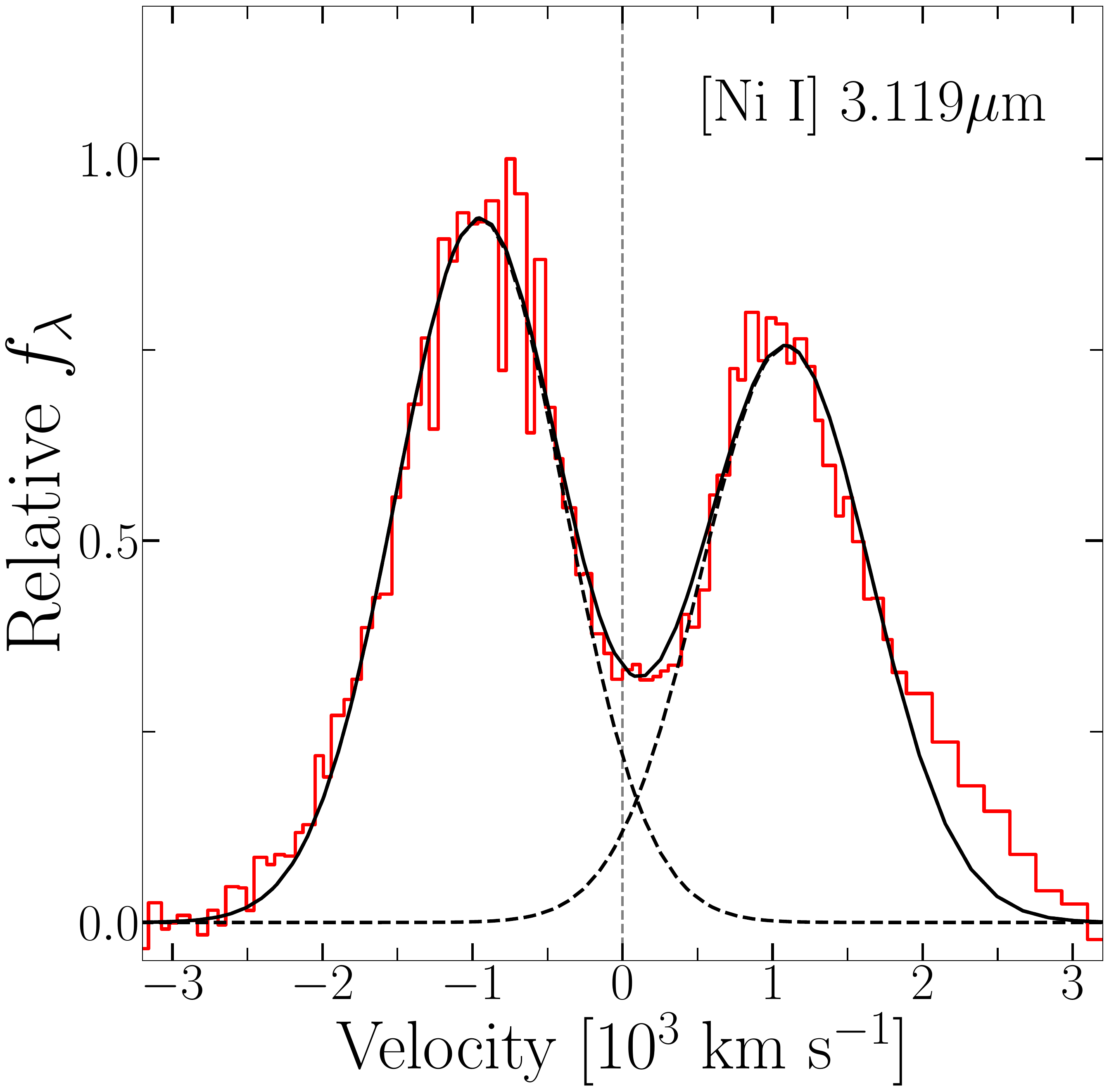}}
\makebox[0.33\textwidth][c]{\hspace*{-2.8mm}\includegraphics[width=0.27\textwidth]{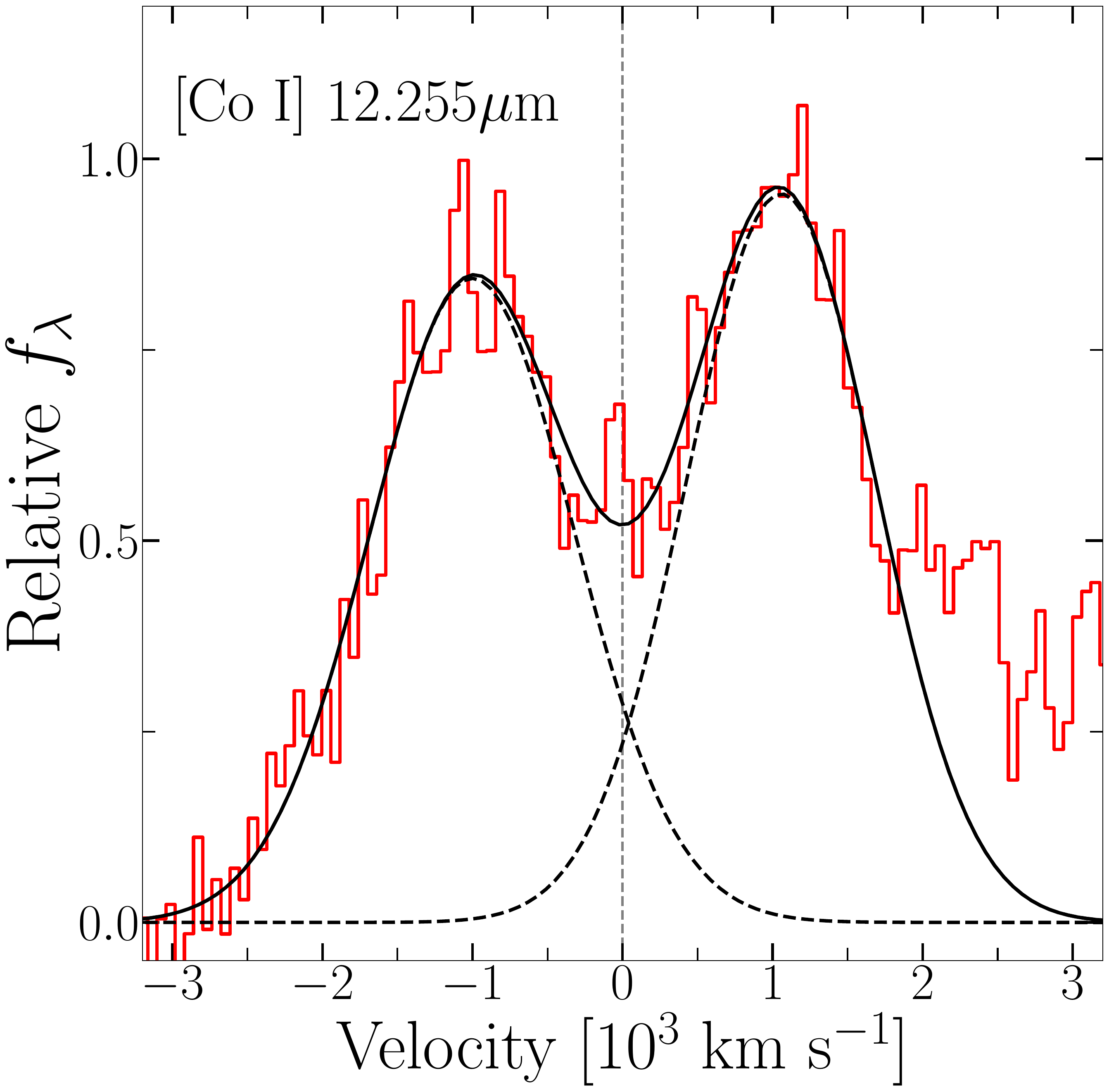}}
\makebox[0.33\textwidth][c]{\hspace*{-2.8mm}\includegraphics[width=0.27\textwidth]{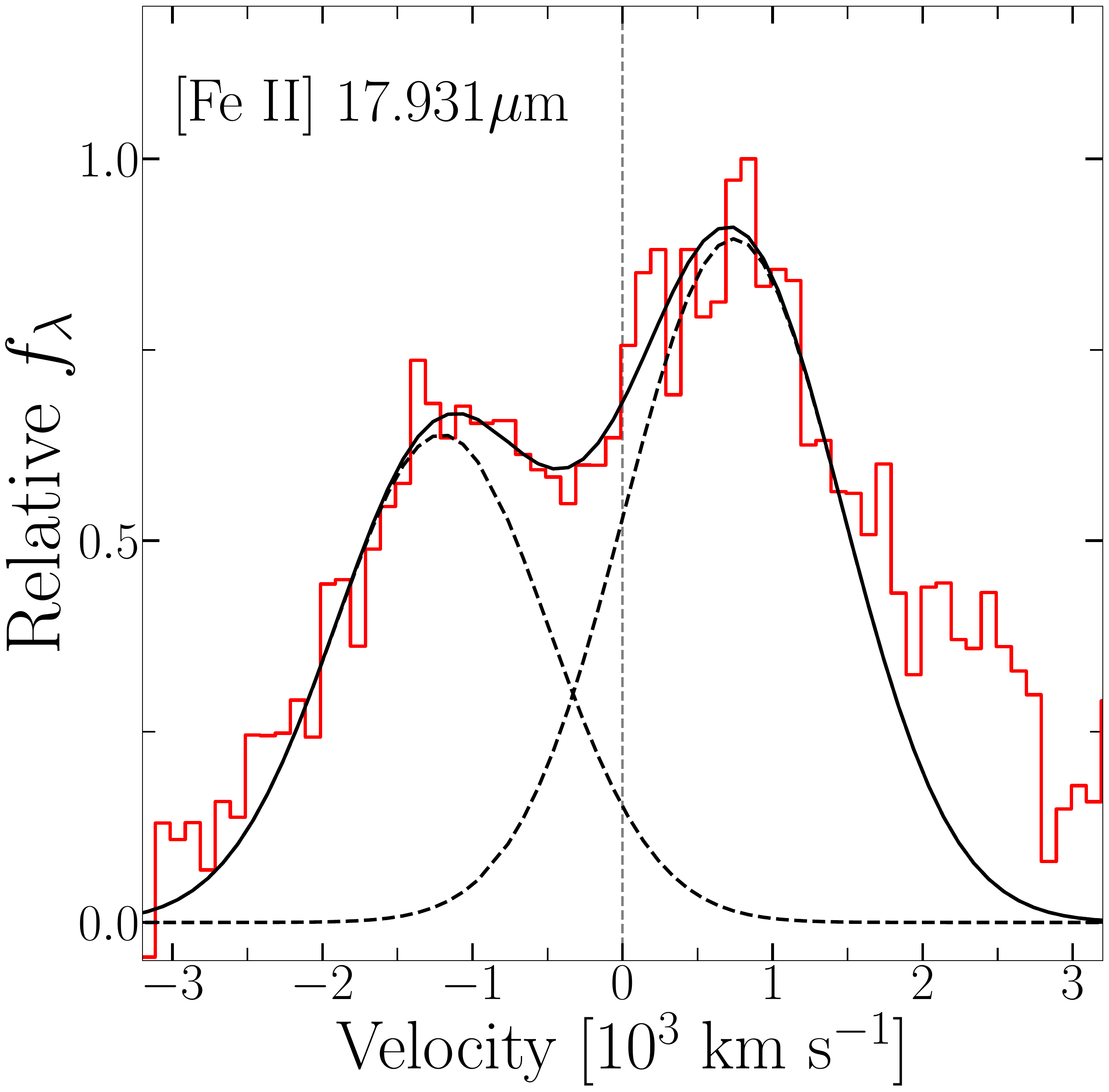}}\\
\caption{Illustrative 2-D velocity-space emissivity model (top) corresponding to a double-Gaussian decomposition of Ni/Co/Fe line profiles (bottom). The two components are placed at ($v_x,v_y$)=($v_{\rm cen},0$) with dispersions $\sigma={\rm FWHM}/2.35$ and amplitudes proportional to the fitted integrated fluxes. Integration of this distribution along the line of sight yields the observed double-peaked spectrum. The orientation within the ($v_x,v_y$) plane is not constrained by the data and is shown for visualization only.   \label{fig:2D} }
\end{figure*}

A similar double-peaked morphology is observed in other IR IGE transitions such as [\ion{Co}{i}] $\lambda$12.255~$\mu$m, [\ion{Fe}{ii}] $\lambda$1.644~$\mu$m and [\ion{Fe}{ii}] $\lambda$17.931~$\mu$m, as well as in the optical [\ion{Fe}{ii}] $\lambda$0.716~$\mu$m line (Fig. \ref{fig:IGEvels}). In Figure \ref{fig:2D}, we visualize the double-peaked [\ion{Ni}{i}] $\lambda$3.119~$\mu$m, [\ion{Co}{i}] $\lambda$12.255~$\mu$m, and [\ion{Fe}{ii}] $\lambda$11.998~$\mu$m profiles by fitting a two Gaussian model and plotting the normalized emissivity in 2-D velocity space. As shown in the contour plots, the blueward [\ion{Ni}{i}] $\lambda$3.119~$\mu$m emission has the highest peak flux, while [\ion{Co}{i}] $\lambda$12.255~$\mu$m and [\ion{Fe}{ii}] $\lambda$11.998~$\mu$m show higher flux in the redward emission. Overall, line profiles probe how much mass/emission is within slabs of thickness $dz$ along the line of sight, where $z = V_r {\rm cos\theta} * t_{\rm SN}$ and there being an infinite number of 3-D configurations (i.e., rearrangements within each slab $dz$) that can reproduce the line profiles.

In order to quantify line asymmetries, we model all double-peaked IGE and IME profiles with two Gaussians and calculate a ``normalized momentum'' parameter $f_{\rm B/R}$, which goes as:

\begin{equation}
    f_B = \frac{L_B |v_{\rm B, peak}|}{L_B |v_{\rm B, peak}| + L_R |v_{\rm R, peak}|} 
\end{equation}

\noindent
where $L_{\rm B/R}$ is amplitude times the full width half maximum (FWHM) of the blue/red Gaussian profile and $v_{\rm B/R, peak}$ is the velocity corresponding to the peak of each Gaussian component. For these resolved IGE emission features, $f_{\rm B/R}$ is approximately equal for both blue and red components of the double Gaussian model. Overall, while the distribution of Ni, Co, and Fe is aspherical and likely concentrated in some part within bipolar lobes, the similarity of $f_{\rm B/R}$ values between the blue- and red-shifted emission components suggests that the explosively synthesized IGEs do not strongly favor either the approaching or receding sides of the ejecta.

\begin{figure*}
\centering
\subfigure{\includegraphics[width=0.49\textwidth]{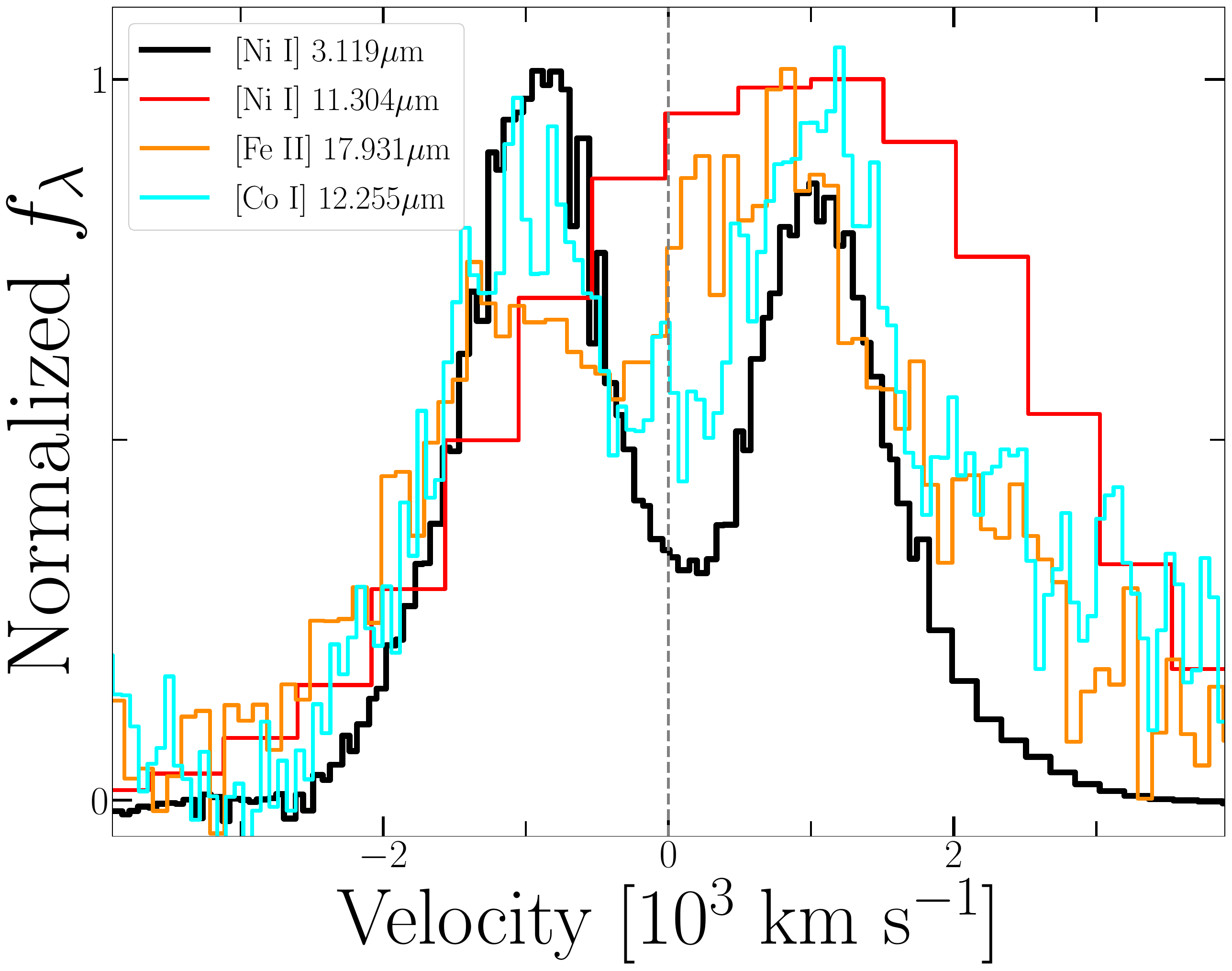}}
\subfigure{\includegraphics[width=0.49\textwidth]{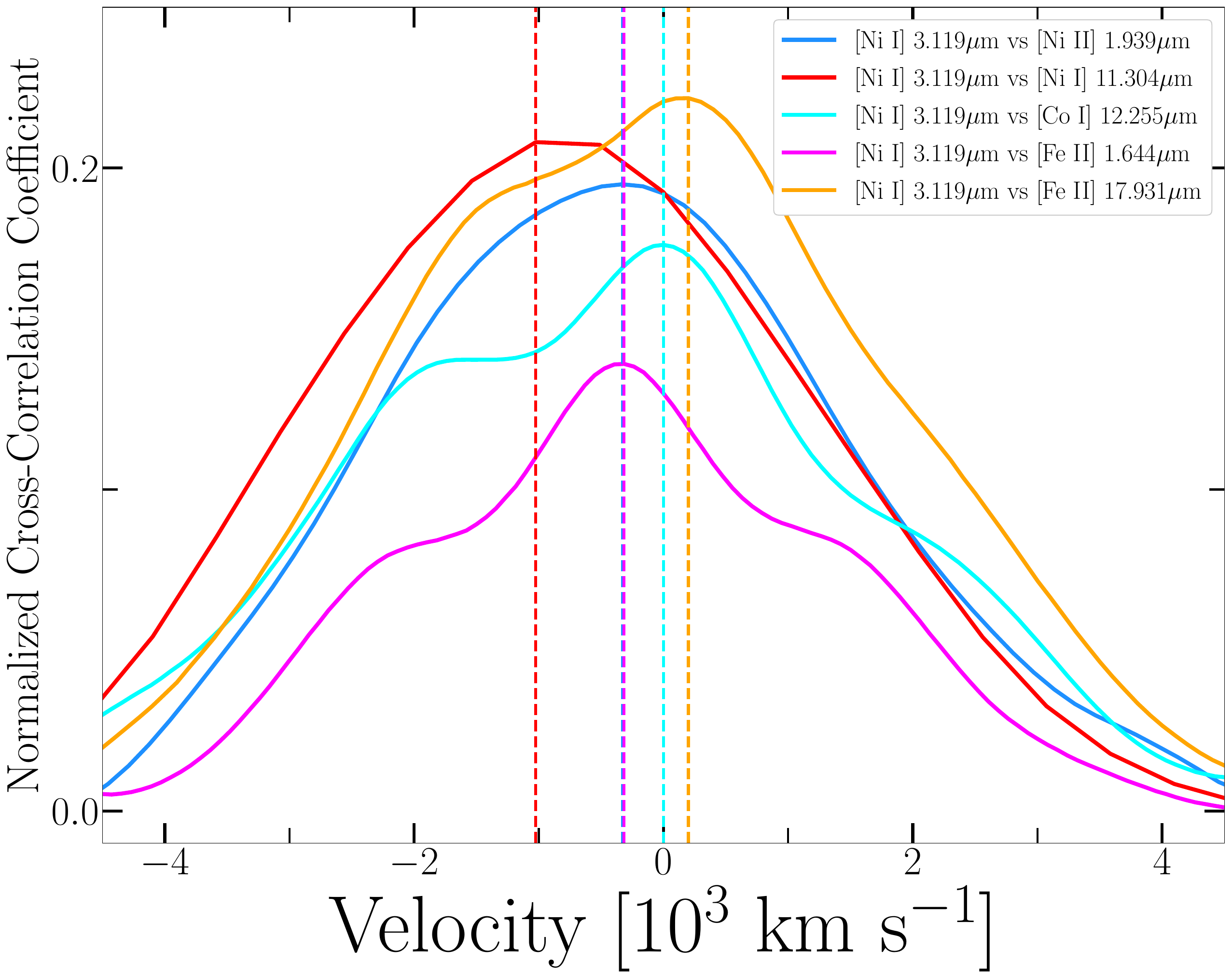}}\\
\subfigure{\includegraphics[width=0.49\textwidth]{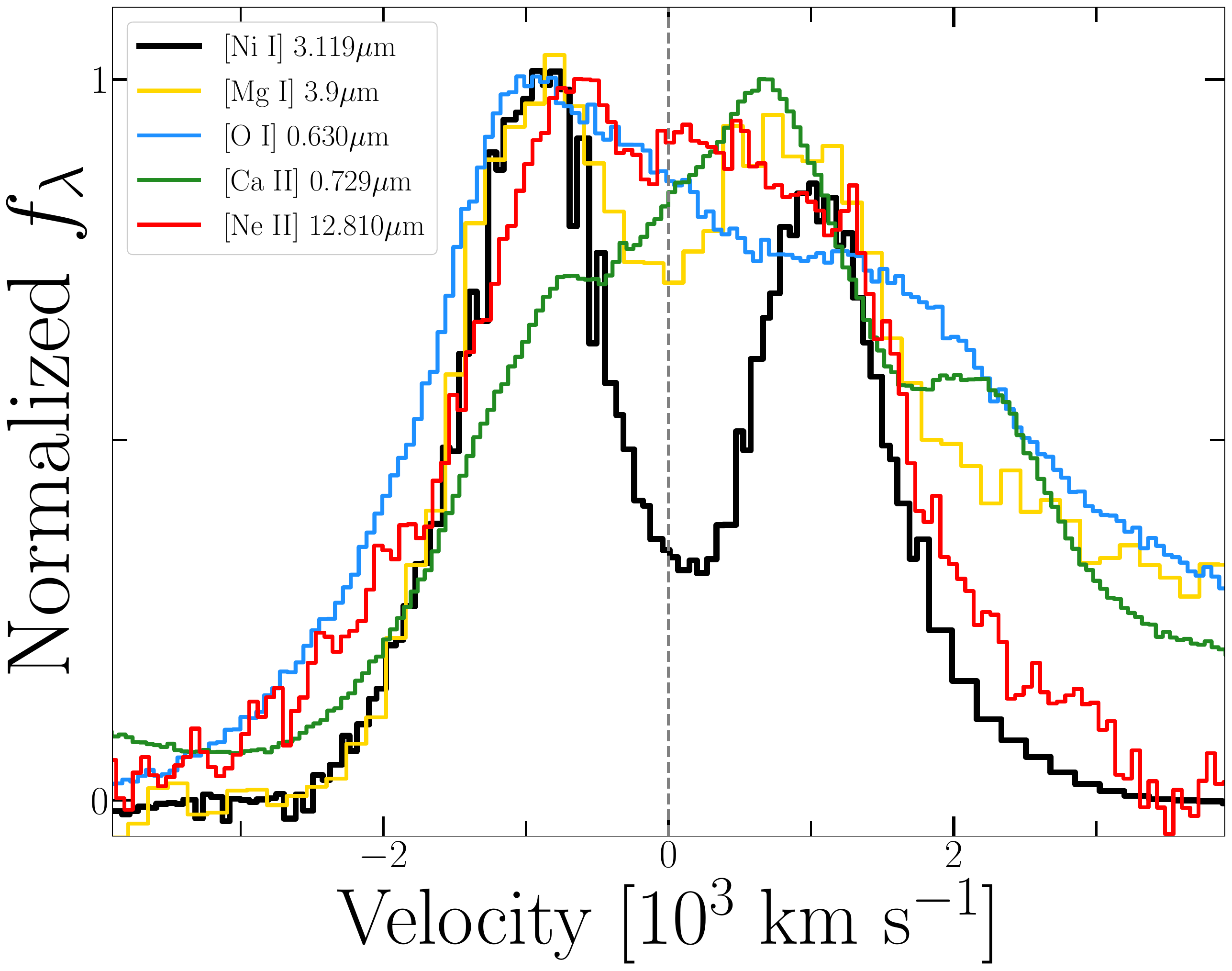}}
\subfigure{\includegraphics[width=0.49\textwidth]{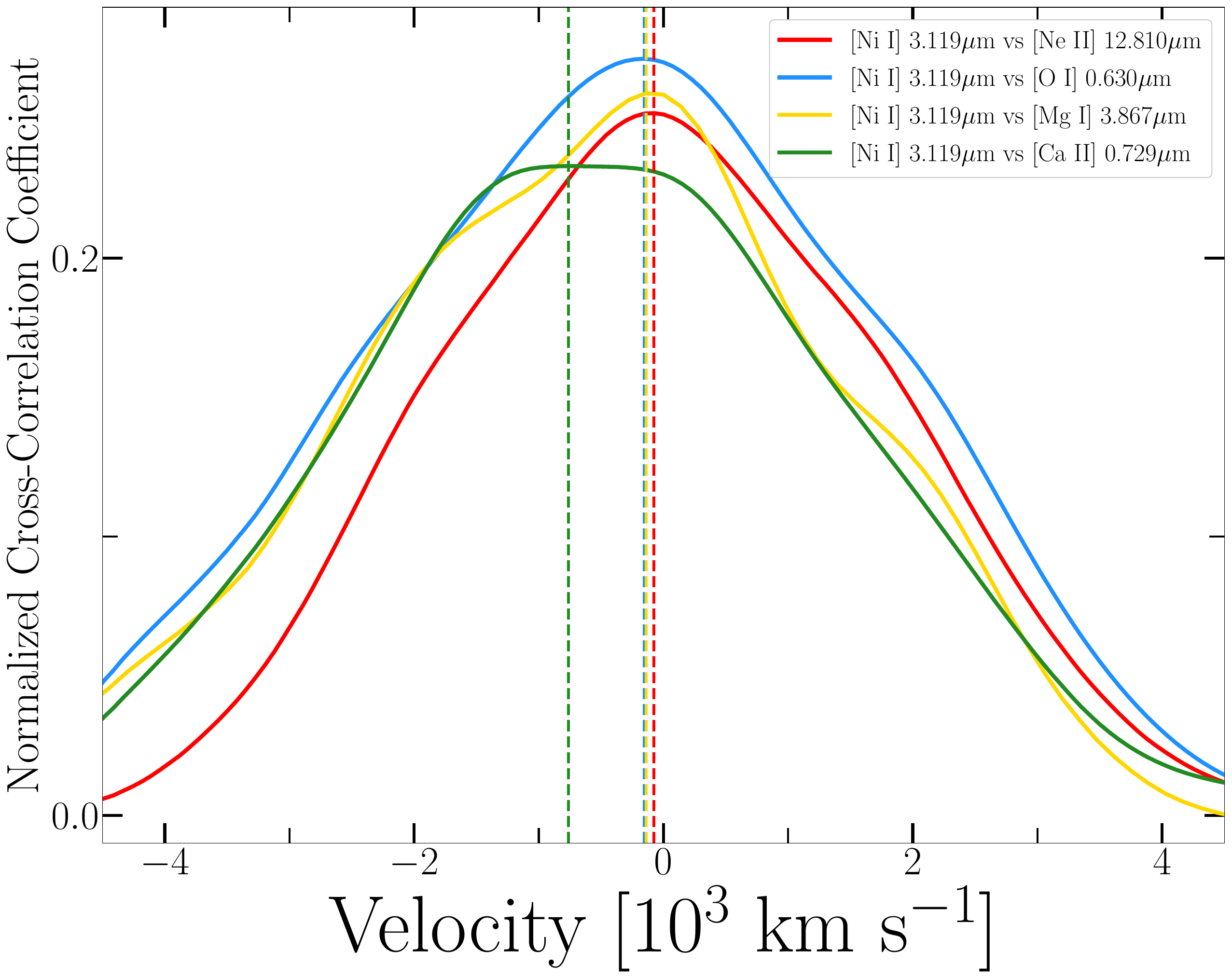}}\\
\caption{ Emission line velocity comparison of [\ion{Ni}{i}] $\lambda$3.119~$\mu$m to IGEs (top left) and IMEs (bottom left). Corresponding cross-correlation coefficient functions (CCFs) shown in top/bottom right panels. All selected IGE and IME emission lines are show a peak correlation with [\ion{Ni}{i}] $\lambda$3.119~$\mu$m at $\sim$0~$\kms$ with the exception of [\ion{Ni}{i}] $\lambda$11.304~$\mu$m and [\ion{Ca}{ii}] $\lambda$0.729~$\mu$m, both of which peak at $\sim$1000~$\kms$ bluewards.  \label{fig:CCF} }
\end{figure*}

We further examine the connections between emission lines by calculating cross-correlation functions (CCFs) for IGE features compared to [\ion{Ni}{i}] $\lambda$3.119~$\mu$m. As shown in Figure \ref{fig:CCF}, other double-peaked IGE lines such as [\ion{Co}{i}] $\lambda$12.255~$\mu$m, [\ion{Fe}{ii}] $\lambda$1.644~$\mu$m, and [\ion{Fe}{ii}] $\lambda$17.931~$\mu$m all show maximum cross-correlation coefficients within $\pm 200~\kms$ of zero velocity. For [\ion{Ni}{ii}] $\lambda$1.939~$\mu$m and [\ion{Ni}{i}] $\lambda$11.304~$\mu$m, both being Gaussian-like, the CCFs have similar peak coefficients as the double-peaked IGE profiles but the velocity at CCF peak is more offset to negative velocities than other features. For example, the [\ion{Ni}{i}] $\lambda$11.304~$\mu$m has a larger blueward spread in its CCF and peaks at $\sim 1000~\kms$, which suggests that while this line and [\ion{Ni}{i}] $\lambda$3.119~$\mu$m may be correlated, they have a distinct spatial distribution within the inner ejecta. 

We repeat the previous exercises for optical and IR IME transitions of O, Ne, Mg, Ar and Ca also presenting their line properties in Table \ref{tab:lines}. Notably, IR IME profiles of [\ion{Mg}{i}] $\lambda$1.502~$\mu$m, [\ion{Mg}{i}] $\lambda$3.867~$\mu$m, and [\ion{Ne}{ii}] $\lambda$12.810~$\mu$m, also show a double-peaked structure, but with a less prominent indentation between peaks than observed in [\ion{Ni}{i}] $\lambda$3.119~$\mu$m. [\ion{Ne}{ii}] $\lambda$12.810~$\mu$m shows a time-dependent morphology as the $\delta t = 266$~day emission line has approximately symmetric blue and red peaks, similar to [\ion{Mg}{i}] $\lambda$3.867~$\mu$m, which then transitions to a more pronounced blue-ward peak by $\delta t = 390$~day. This asymmetric, blue-dominant profile matches that of [\ion{Mg}{i}] $\lambda$1.502~$\mu$m and [\ion{O}{i}] $\lambda$0.630~$\mu$m at similar phases. The double-peaked structure of [\ion{Mg}{i}] $\lambda$1.502~$\mu$m and [\ion{Fe}{ii}] $\lambda$1.644~$\mu$m was also noted in the analysis of NIR nebular spectra of SN~2024ggi by \cite{Hueichap26}. Furthermore, the resulting values of $f_{\rm B} \approx 0.3$ and $f_{\rm R} \approx 0.7$ for both [\ion{Mg}{i}] lines suggest larger asymmetry than [\ion{Ne}{ii}] $\lambda$12.810~$\mu$m and double-peaked IGE profiles. Despite the detection of molecular emission (e.g., see \citealt{Dessart24ggi, Mera26}), this asymmetry is unlikely related to significant dust attenuation given the symmetry of the H profiles, which are well-reproduced by the s15p2 model (Fig. \ref{fig:all}). Finally, we compute CCFs for select IMEs profiles compared to [\ion{Ni}{i}] $\lambda$3.119~$\mu$m (Fig. \ref{fig:CCF}) and find a correlation peak near $0~\kms$. [\ion{Ca}{ii}] $\lambda$0.729~$\mu$m does show a skewed CCF that peaks at $\sim -800~\kms$, which is consistent with overall shape of the profile compared to other IMEs. 


\subsubsection{Constraining Density Profiles Per Species}\label{subsubsection:density}

\begin{figure*}
\centering
\subfigure{\includegraphics[width=0.33\textwidth]{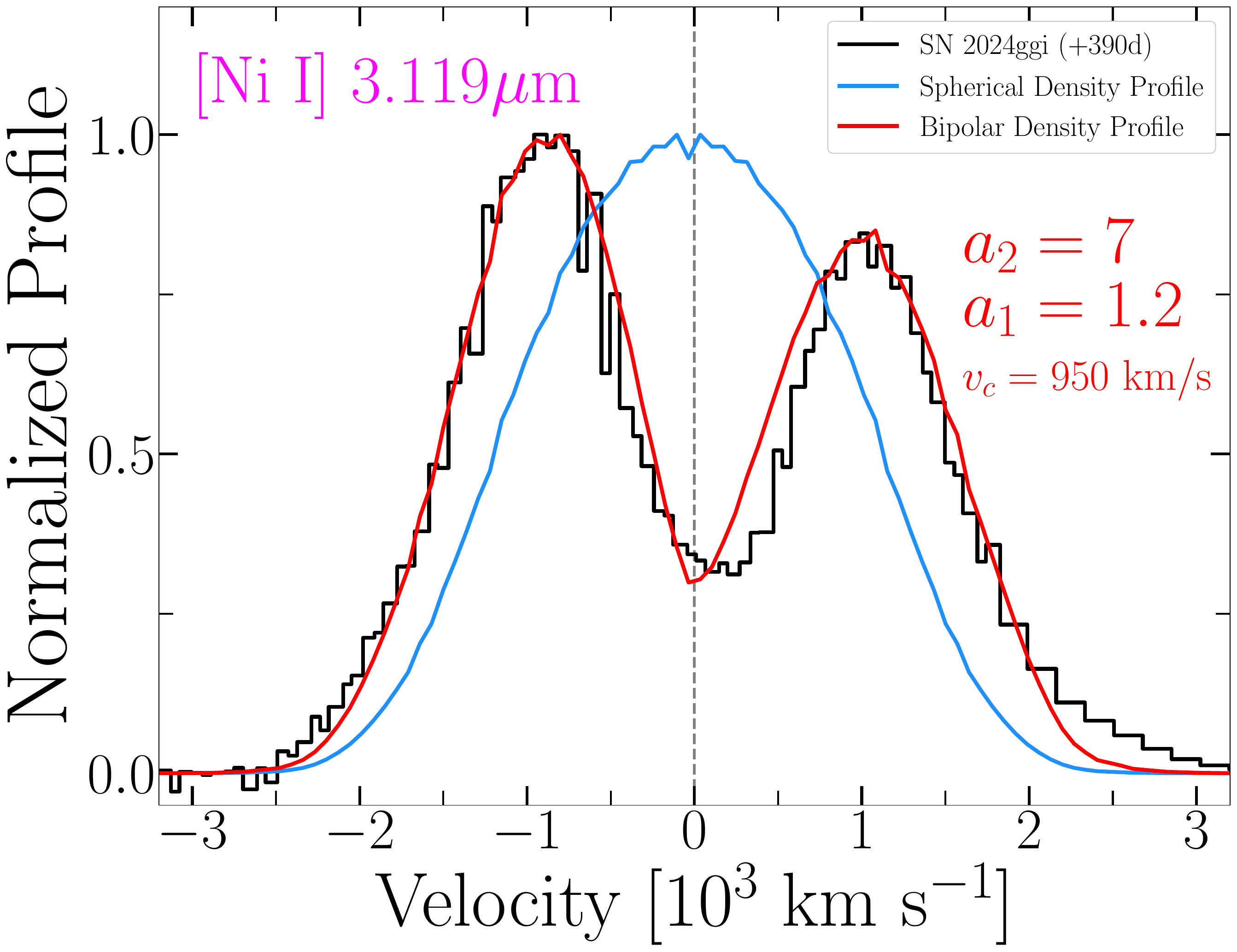}}
\subfigure{\includegraphics[width=0.33\textwidth]{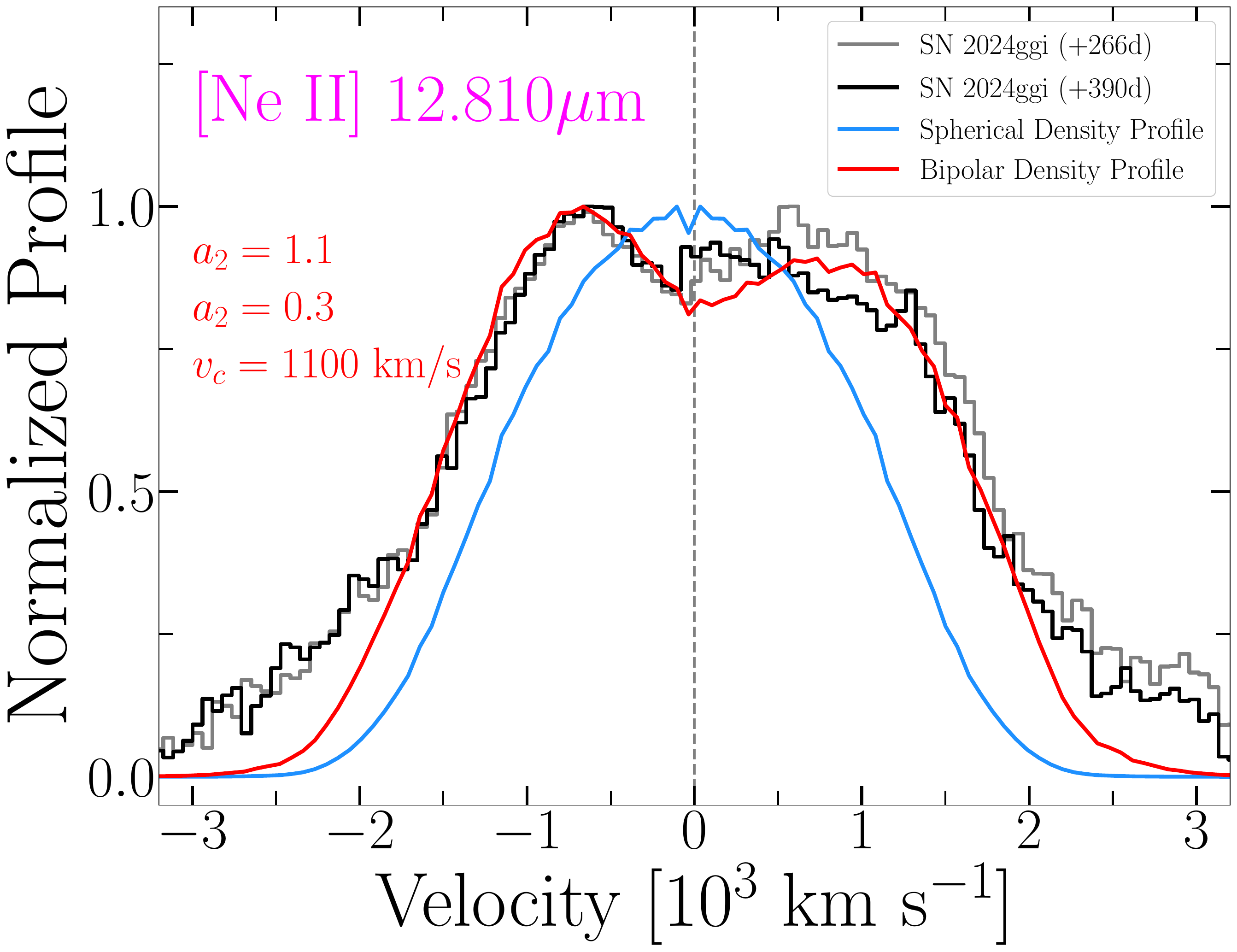}}
\subfigure{\includegraphics[width=0.33\textwidth]{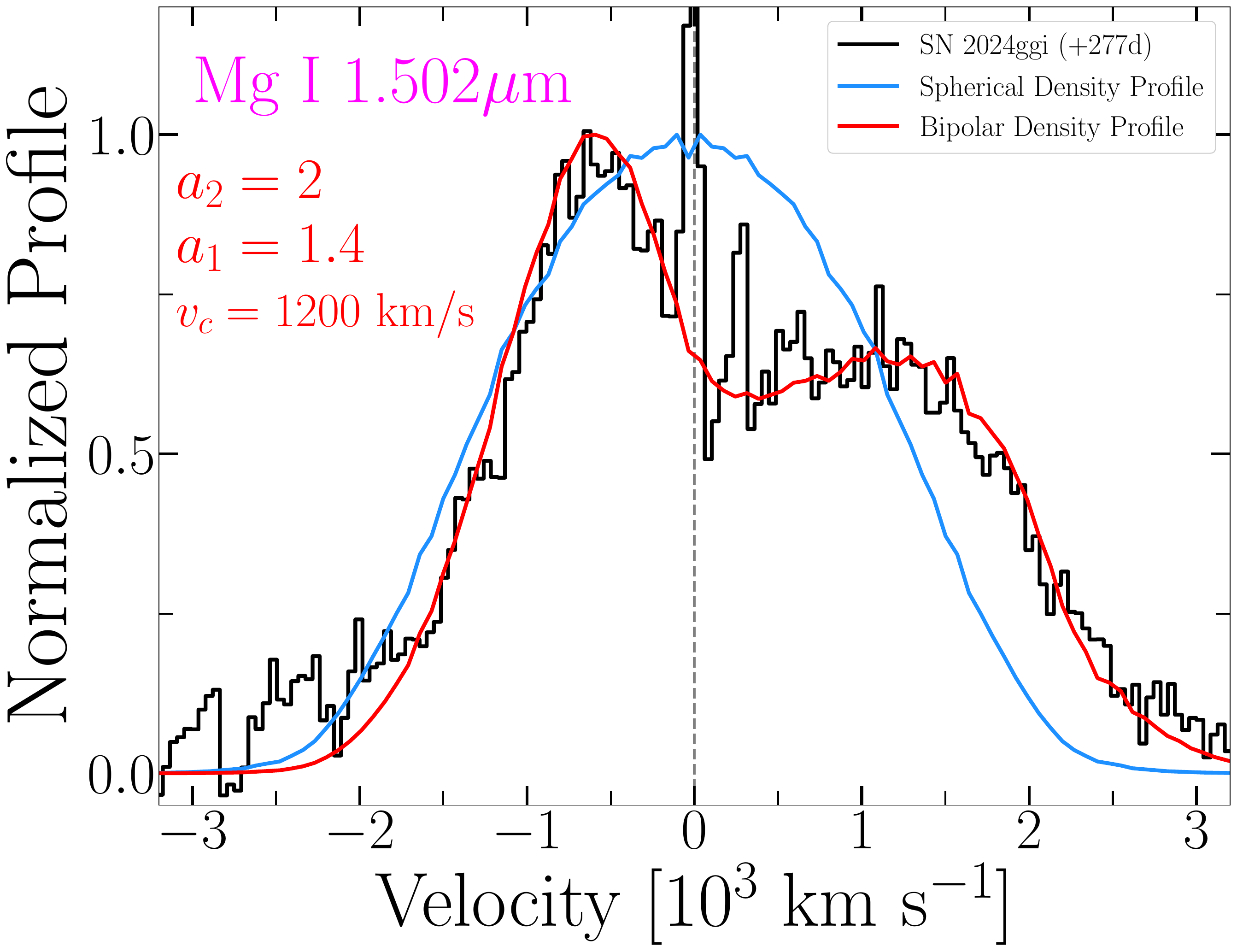}}
\caption{Emission line provide versus Doppler velocities of [\ion{Ni}{i}] $\lambda$3.1199~$\mu$m ({\it left}), [\ion{Ne}{ii}] $\lambda$12.8101~$\mu$m ({\it middle}), and [\ion{Mg}{i}] $\lambda$1.50202~$\mu$m ({\it right}), all normalized to peak flux. SN~2024ggi shown in black compared to model density profiles with spherical (blue) and bipolar (red) density profiles, as discussed in \S\ref{subsubsection:density}. [\ion{Ni}{i}] requires the most significant polar enhancement ($a_2 = 7$) compared to IME profiles. \label{fig:densityprofiles} }
\end{figure*}

To quantify the degree of asymmetry required to reproduce observed line profiles, we construct simple parametric abundance distributions with the s15p2 \citep{Dessart21} model ejecta structure at $\delta t = 400$~days. Because the nebular IR line profiles largely trace the mass distribution in line-of-sight velocity space, these experiments provide an intuitive mapping between distribution of elements in the ejecta and the observed spectral morphology (e.g., see \S\ref{subsec:cmfgen}). In all cases, the species abundance distribution $X_{\rm species}$ is allowed to vary as a function of both radial velocity and polar angle while preserving the same total species mass. The 2-D distribution of the emitting material is parameterized as:

\begin{equation}
X(v,\theta)=
\begin{cases}
X_0, & v < v_c(\theta), \\
X_0 \exp\left[-\left(\frac{v-v_c(\theta)}{\Delta v(\theta)}\right)^2\right], & v \ge v_c(\theta),
\end{cases}
\end{equation}

\noindent
where $v_c(\theta)$ sets the characteristic velocity extent of the emitting volume for a given species and $\Delta v(\theta)$ controls the sharpness of the abundance decline at larger velocities. Smaller values of $\Delta v$ produce thinner, shell-like distributions and sharper double-peaked profiles, while larger values yield more extended distributions with smoother/broader emission profiles. A spherically symmetric distribution is represented by $X(v,\theta)=X(v)$, which produces a centrally-peaked Gaussian profile as shown in Figure~\ref{fig:densityprofiles}. To generate a variety of bipolar ejecta morphologies, we allowed both the characteristic velocity extent $v_c(\theta)$ and abundance normalization $X(\theta)$ to vary with polar angle according to:

\begin{equation}
v_c(\theta) = v_{c,0}\left[1+a_{v,1}\cos\theta + a_{v,2}\cos^2\theta \right],
\end{equation}

\begin{equation}
\Delta v(\theta) = \Delta v_0\left[1+a_{\Delta v,1}\cos\theta+a_{\Delta v,2}\cos^2\theta\right],
\end{equation}

\begin{equation}
X(\theta)\propto 1+a_1\cos\theta+a_2\cos^2\theta .
\end{equation}

\noindent
The coefficients $a_{v,2}$ and $a_{\Delta v,2}$ control the degree to which the emitting material extends to larger velocities along the poles, while $a_{v,1}$ and $a_{\Delta v,1}$ introduce front/back asymmetries in the radial velocity distribution. Separately, $a_2$ controls the strength of polar abundance enhancement and $a_1$ modifies the relative emissivity of approaching and receding hemispheres. This prescription naturally produces symmetric double-peaked profiles while allowing blue and red peaks to differ in both amplitude and velocity extent. Increasing $a_2$ strengthens the peak height and suppresses contribution from low projected velocities, whereas increasing $a_{v,2}$ shifts more emitting material to larger projected velocities and increases the peak separation. Positive values of $a_1$ enhance the approaching hemisphere and produce stronger blue-shifted emission, while positive values of $a_{v,1}$ increase the velocity extent of one hemisphere relative to the other.

In Figure \ref{fig:densityprofiles}, we illustrate how these parameters can be modified to match the emission line morphology of specific IGE and IME profiles observed in SN~2024ggi. In order to reproduce the double-peaked profile of [\ion{Ni}{i}] $\lambda$3.119~$\mu$m, the model requires a characteristic velocity of $v_c = 950~\kms$, $\Delta v = 850~\kms$, a polar enhancement of $a_2 = 7$, and a blueward enhancement of $a_2 = 1.2$. The similar double-peaked IGE profiles of Ni, Co and Fe (Fig. \ref{fig:IGEvels}) suggests that these ions also require increased density in polar regions with $a_2 \approx 2-4$, but may have variations in $a_1$ based on the line morphology e.g., [\ion{Co}{i}] $\lambda$12.255~$\mu$m and [\ion{Fe}{ii}] $\lambda$17.931~$\mu$m need redward enhancement to match their profiles. 

We further explore this modeling approach for the double-peaked and asymmetric profiles of IMEs [\ion{Ne}{ii}] $\lambda$12.810~$\mu$m and [\ion{Mg}{i}] $\lambda$1.502~$\mu$m. For Ne, we find that values of $a_2 = 1.1$, $a_2 = 0.3$, $v_c = 1100~\kms$, and $\Delta v = 1000~\kms$ can reproduce the profile morphology, suggesting minor polar enhancement with increased velocity in the receding material. However, the extended wings of the profile are not captured by the model so there is likely emitting material at faster velocities. For Mg, the profile is highly asymmetric and requires model parameters of $a_2 = 2$, $a_1 = 1.4$,  $v_c = 1000~\kms$, and $\Delta v = 1000~\kms$. Overall, this material has slightly increased polar densities and significant blueward enhancement as well as an increased velocity in the receding material, similar to Ne. However, we note that there is a contribution from overlapping Fe transitions on the redward side of the profile, suggesting that the Mg blue/red asymmetry could be even more extreme. From this, we can conclude that most emission is spread over a similar extent, irrespective of species, but variations in line morphology implies diverse levels of asymmetry across element distributions.

\subsubsection{Ion Mass Estimation}

In Table \ref{tab:lines}, we present line luminosities for all prominent IGE and IME ions in the optical-IR spectra of SN~2024ggi at both late-time epochs. All line luminosities were derived after subtracting continuum emission and we integrate the line flux from Gaussian model fits for all lines that are not fully isolated (e.g., [\ion{Mg}{i}] $\lambda$3.867~$\mu$m). We then attempt to derive stable Ni masses per ion by first assuming an LTE approximation for level populations set by a Boltzmann distribution and optically-thin ejecta emission \citep{osterbrock06, Jerkstrand17}. Here, the number of ions in the upper level goes as:

\begin{equation}
    N_u = \frac{L_{ul}}{A_{ul} \Delta E}
\end{equation}

\noindent
where $L_{ul}$ is ion luminosity, $A_{ul}$ is Einstein A value, and $\Delta E$ is level excitation energy. Assuming a two level approximation, the total mass of a given ion in its upper level goes as: 

\begin{equation}
    M_{\rm ion} = M_u
\left[
1 + \frac{g_l}{g_u} e^{\Delta E / kT}
\right]
\end{equation}

\noindent
where $M_u = N_u m_{\rm atom}$, $T$ is gas temperature, $g$ are the level degeneracies and $k$ is the Boltzmann constant. Adopting a gas temperature of $T = 5000$~K, we find a [\ion{Ni}{ii}] upper-level mass of $M_{\rm Ni \ II} = (1.9 \pm 0.1) \times 10^{-3}~\Msun$ using the $\lambda$6.634~$\mu$m transition properties ($g_l / g_u = 3/2$) and line luminosity at $\delta t = 390$~days. We repeat this calculation for other IGE transitions to find a [\ion{Ni}{i}] $\lambda$3.119~$\mu$m ($g_l / g_u = 7/5$) upper-level mass of $M_{\rm Ni \ I} = (1.4 \pm 0.1) \times 10^{-4}~\Msun$, a [\ion{Co}{i}] $\lambda$12.255~$\mu$m ($g_l / g_u = 5/4$) upper-level mass of $M_{\rm Co \ I} = (5.5 \pm 0.3) \times 10^{-4}~\Msun$, and a [\ion{Co}{ii}] $\lambda$10.520~$\mu$m ($g_l / g_u = 9/7$) upper-level mass of $M_{\rm Co \ II} = (9.0 \pm 0.5) \times 10^{-3}~\Msun$. We note that adopting a lower gas temperature of $T = 3000$~K increases the upper-level mass by a factor of $\leq 2$.

In order to estimate the total Ni and Co mass in SN~2024ggi, we need to estimate the ionization fraction, which goes as $f_{\rm Ni \ II} = M_{\rm Ni \ II} / (M_{\rm Ni \ I} + M_{\rm Ni \ II})$ for $^{58}$Ni using the calculated upper-level masses. With $f_{\rm Ni \ II} = 0.93$ and $f_{\rm Co \ II} = 0.62$, we calculate total Ni and Co masses of $M_{\rm Ni} = M_{\rm Ni \ II} / f_{\rm Ni \ II} = (2.1 \pm 0.1) \times 10^{-3}~\Msun$ and $M_{\rm Co} = M_{\rm Co \ II} / f_{\rm Co \ II} = (1.4 \pm 0.1) \times 10^{-3}~\Msun$. For a $^{56}$Ni mass of $0.05~\Msun$ \citep{Dessart24ggi, Ferrari25}, we find a radioactive to stable Ni mass ratio of $\sim 24$, which is consistent with theoretical predictions used in core-collapse SN simulations (e.g., \citealt{Jerkstrand15Fe, Sukhbold16, Dessart21}). We check the accuracy of this LTE calculation by calculating the stable Ni mass using emission line luminosities in the s15p2 model at 400~days, which has an input $^{58}$Ni mass of $3.7\times 10^{-3}~\Msun$ \citep{Dessart25IR}. From the [\ion{Ni}{i}] $\lambda$3.119~$\mu$m and [\ion{Ni}{ii}] $\lambda$6.634~$\mu$m luminosities, we calculate $M_{\rm Ni \ I} = 2.8 \times 10^{-4}~\Msun$, $M_{\rm Ni \ II} = 1.5 \times 10^{-3}~\Msun$, and $f_{\rm Ni \ II} = 0.84$. This then yields a total stable Ni mass of $M_{\rm Ni} = 1.8 \times 10^{-3}~\Msun$, which is a factor of $\sim 2.1$ smaller than the model input $^{58}$Ni mass. Consequently, the stable Ni mass calculated for SN~2024ggi is likely a lower limit on the true $^{58}$Ni mass in the ejecta. Furthermore, at $\delta t = 400$~days, the s15p2 model has $M(^{56}\rm Co) = 1.8 \times 10^{-3}~\Msun$ and $M(^{59}\rm Co) = 1.8 \times 10^{-4}~\Msun$, indicating that the Co mass estimated for SN~2024ggi is derived from both stable and radioactive Co luminosities and likely an over-estimate.

As noted in numerous works on SN spectral modeling, inner ejecta conditions at nebular phases requires non-LTE calculations for the radiative transfer, which include thermal or non-thermal processes, photo-excitation and recombination, all of which can contribute to populating various energy levels (e.g., see \citealt{dessart11, Jerkstrand17, Dessart21, Jerkstrand25}). Overall, the regime in which non-LTE effects should be considered is the low density limit, $n_e << n_{\rm crit}$, where critical density goes as $n_{\rm crit} = A_{ul} / q_{ul}$. Conversely, LTE approximations are appropriate in the high density limit when $n_e >> n_{\rm crit}$. Here, $n_e$ can be written as:

\begin{equation}
    n_e = \frac{3\,M_{\rm ion}}{4\pi (v t)^3 f\,\mu_e m_p}
\end{equation}

\noindent
where $M_{\rm ion}$ is ion mass, $v$ is characteristic expansion velocity of that ion, $t$ is time since first light, $f$ is volume filling factor, $\mu_e$ is mean mass per free electron (we adopt $\mu_e = 2$ for the Fe-rich inner ejecta), and $m_p$ is proton mass. For [\ion{Ni}{ii}], we adopt $v = 2000~\kms$ based on the velocity spread observed in the line profile (Fig. \ref{fig:IGEvels}), $t = 390$~days, and $f = 0.15$ based on \cite{Jerkstrand12} noting however that the filling factor is highly uncertain. For these values and the ion mass calculated above, we find $n_e \approx 6 \times 10^6$~cm$^{-3}$ for [\ion{Ni}{ii}] $\lambda$6.634~$\mu$m, which is only a factor of $\sim 2$ larger than the $n_{\rm crit} = 2.7 \times 10^{6}$~cm$^{-3}$ value shown in Table \ref{tab:lines}. However, for $f = 0.5$, $n_e$ would have a smaller value of $n_e \approx 2 \times 10^6$~cm$^{-3}$, which is less than $n_{\rm crit}$, and highlights that ion mass estimates are also uncertain to explosion asymmetry and viewing angle effects. Consequently, this does not reflect a purely non-LTE regime but LTE assumptions are also likely not appropriate, indicating that the ion mass should be treated as a lower limit.  

\begin{figure*}
\centering
\subfigure{\includegraphics[width=0.49\textwidth]{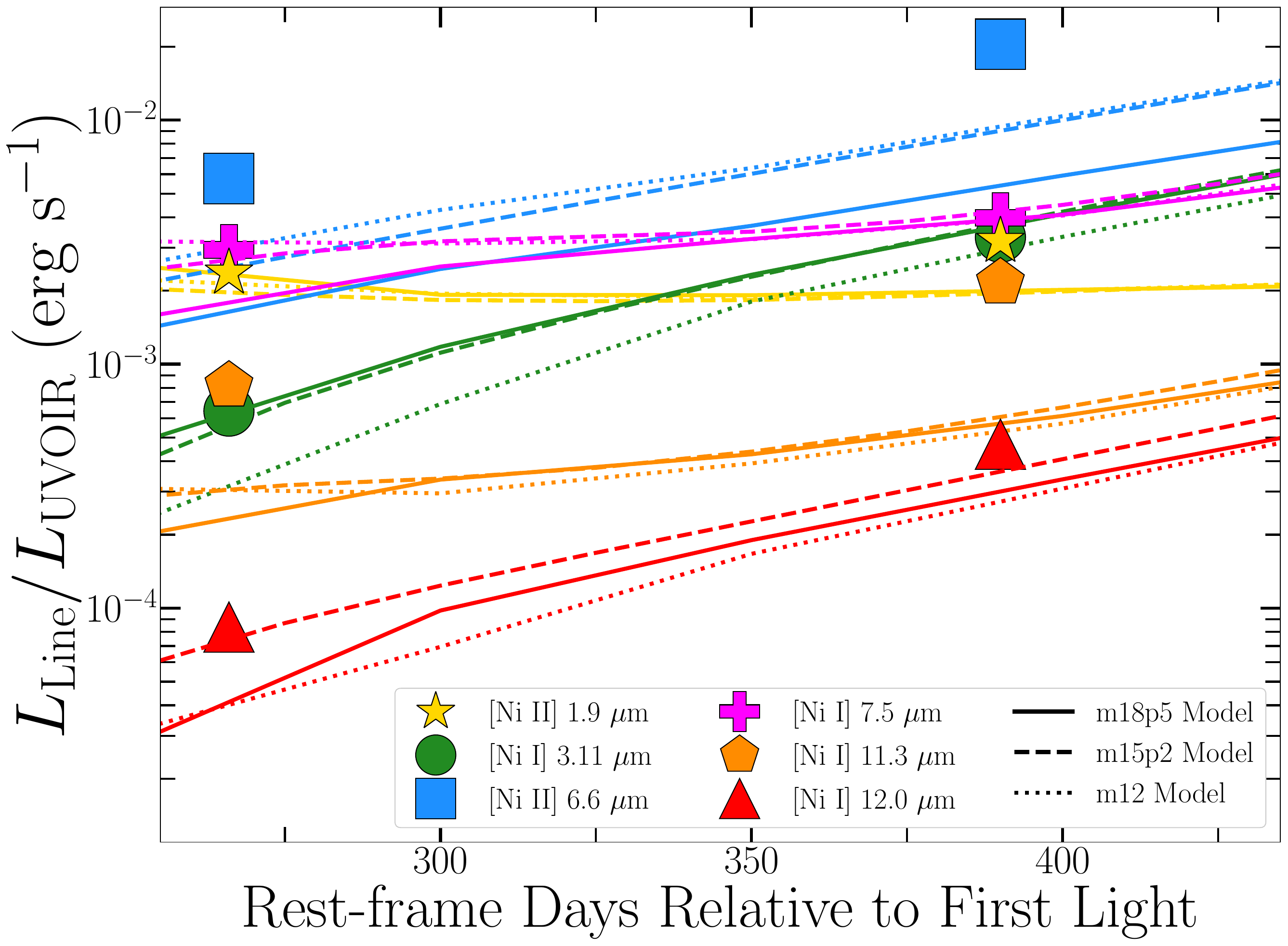}}
\subfigure{\includegraphics[width=0.49\textwidth]{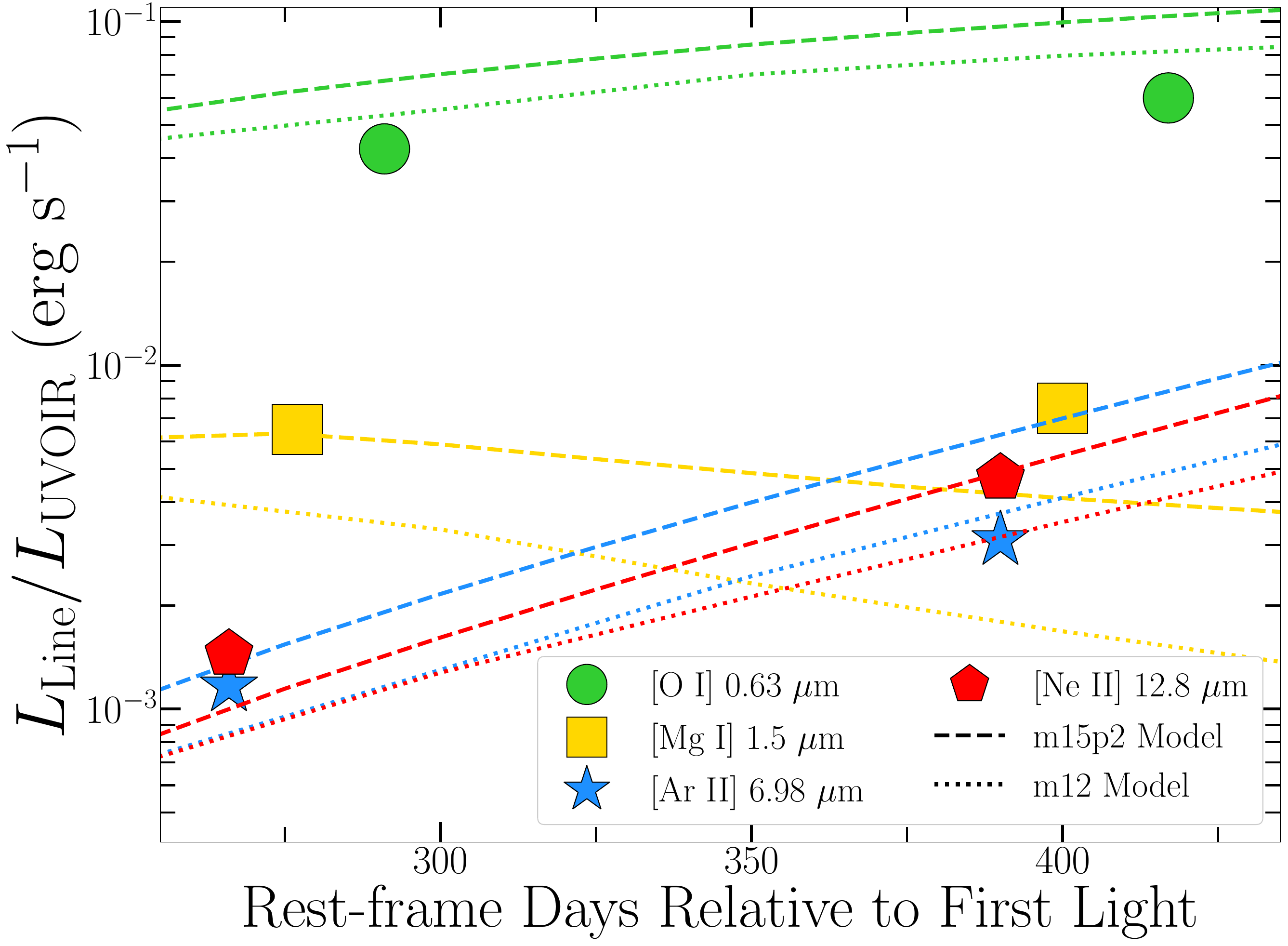}}
\caption{{\it Left:} Measured, multi-epoch emission line luminosities of [\ion{Ni}{ii}] $\lambda$1.939~$\mu$m (yellow stars), [\ion{Ni}{i}] $\lambda$3.119~$\mu$m (green circles), [\ion{Ni}{ii}] $\lambda$6.634~$\mu$m (blue squares), [\ion{Ni}{i}] $\lambda$7.505~$\mu$m (magenta plus signs), [\ion{Ni}{i}] $\lambda$11.304~$\mu$m (orange polygons), and [\ion{Ni}{i}] $\lambda$11.998~$\mu$m (red triangles) in the nebular IR spectra of SN~2024ggi compared to s12 ($M_{\rm ZAMS} = 12~\Msun$), s15p2 ($M_{\rm ZAMS} = 15.2~\Msun$), and s18p5 ($M_{\rm ZAMS} = 18.5~\Msun$) models shown as dotted, dashed and solid lines, respectively. All line luminosities have been normalized by the total spectral flux between $0.35 - 15~\mu$m in order to best represent the absorbed radioactive decay power. {\it Right:} Measured, multi-epoch emission line luminosities of [\ion{O}{i}] $\lambda$0.630~$\mu$m (green circles), [\ion{Mg}{i}] $\lambda$1.502~$\mu$m (yellow squares), [\ion{Ar}{ii}] $\lambda$6.983~$\mu$m (blue stars), and [\ion{Ne}{ii}] $\lambda$12.810~$\mu$m (red polygons). \label{fig:lineflux} }
\end{figure*}

\subsection{Progenitor ZAMS Mass Constraints}

A common practice in SN~II nebular spectra analysis is to compare the line luminosities of forbidden optical emission lines such as [\ion{O}{i}] and [\ion{Ca}{ii}] to grids of model spectra (e.g., \citealt{Jerkstrand14, Dessart21}) in order to derive a corresponding progenitor zero age main sequence (ZAMS) mass (e.g., see \citealt{Maguire12, Fang25neb, wjg25}). Using the [\ion{O}{i}] luminosity in nebular spectroscopy of SN~2024ggi, \cite{Ferrari25} estimate $M_{\rm ZAMS} = 12 - 15~\Msun$, while \cite{Hueichap26} infer $M_{\rm ZAMS} \approx 14~\Msun$. These estimates are consistent with the pre-explosion estimate of $M_{\rm ZAMS} = 13 \pm 1~\Msun$ from \cite{Xiang24} as well as progenitor masses inferred from bolometric light curve modeling e.g., $\sim 15~\Msun$ from \cite{Ertini25} and $11 \pm 1~\Msun$ from \cite{Aryan25}. In Figure \ref{fig:lineflux}, we test additional emission line fluxes in both the optical and IR to SN~II nebular spectra models from \cite{Dessart25IR} given the global consistency found between the SN~2024ggi optical/IR nebular spectra and the s15p2 ($M_{\rm ZAMS} = 15.2~\Msun$) \cmfgen\ model found by \cite{Dessart24ggi}. 

In the left panel of Figure \ref{fig:lineflux}, we compare prominent near- and mid-IR [\ion{Ni}{i/ii}] line luminosities of SN~2024ggi at both late-time epochs to \cmfgen\ models for 12, 15.2 and 18.5~$\Msun$ progenitor ZAMS masses. We normalize all line luminosities by dividing out the total flux between $0.35-15~\mu$m, which represents the emergent luminosity from the absorbed radioactive decay power. We find that [\ion{Ni}{ii}] $\lambda$1.939~$\mu$m and [\ion{Ni}{i}] $\lambda11.998$~$\mu$m lines are most consistent with the $M_{\rm ZAMS} = 15.2~\Msun$ model, while the [\ion{Ni}{i}] $\lambda7.505$~$\mu$m luminosity is best matched to the $M_{\rm ZAMS} = 12~\Msun$ model and [\ion{Ni}{i}] $\lambda3.11$~$\mu$m lies in between 12 and 15~$\Msun$. Both the [\ion{Ni}{ii}] $\lambda6.634$~$\mu$m and [\ion{Ni}{i}] $\lambda7.505$~$\mu$m luminosities in SN~2024ggi are larger than all three models, and it is worth noting that not all Ni line luminosities increase with larger $M_{\rm ZAMS}$ e.g., the s18p5 model has a lower [\ion{Ni}{ii}] $\lambda6.634$~$\mu$m luminosity than the s15p2 model. For IME lines, [\ion{Ne}{ii}] $\lambda12.810$~$\mu$m, [\ion{Mg}{i}] $\lambda1.502$~$\mu$m and [\ion{Ar}{ii}] $\lambda6.983$~$\mu$m are bracketed by 12 and 15~$\Msun$ models, which is also consistent with [\ion{O}{i}] $\lambda0.630$~$\mu$m luminosity constraints. Overall, IR line emission can be used as additional probes of late-stage burning and constraining the final progenitor core mass in SNe~II. 

\begin{figure*}
\centering
\subfigure{\includegraphics[width=0.33\textwidth]{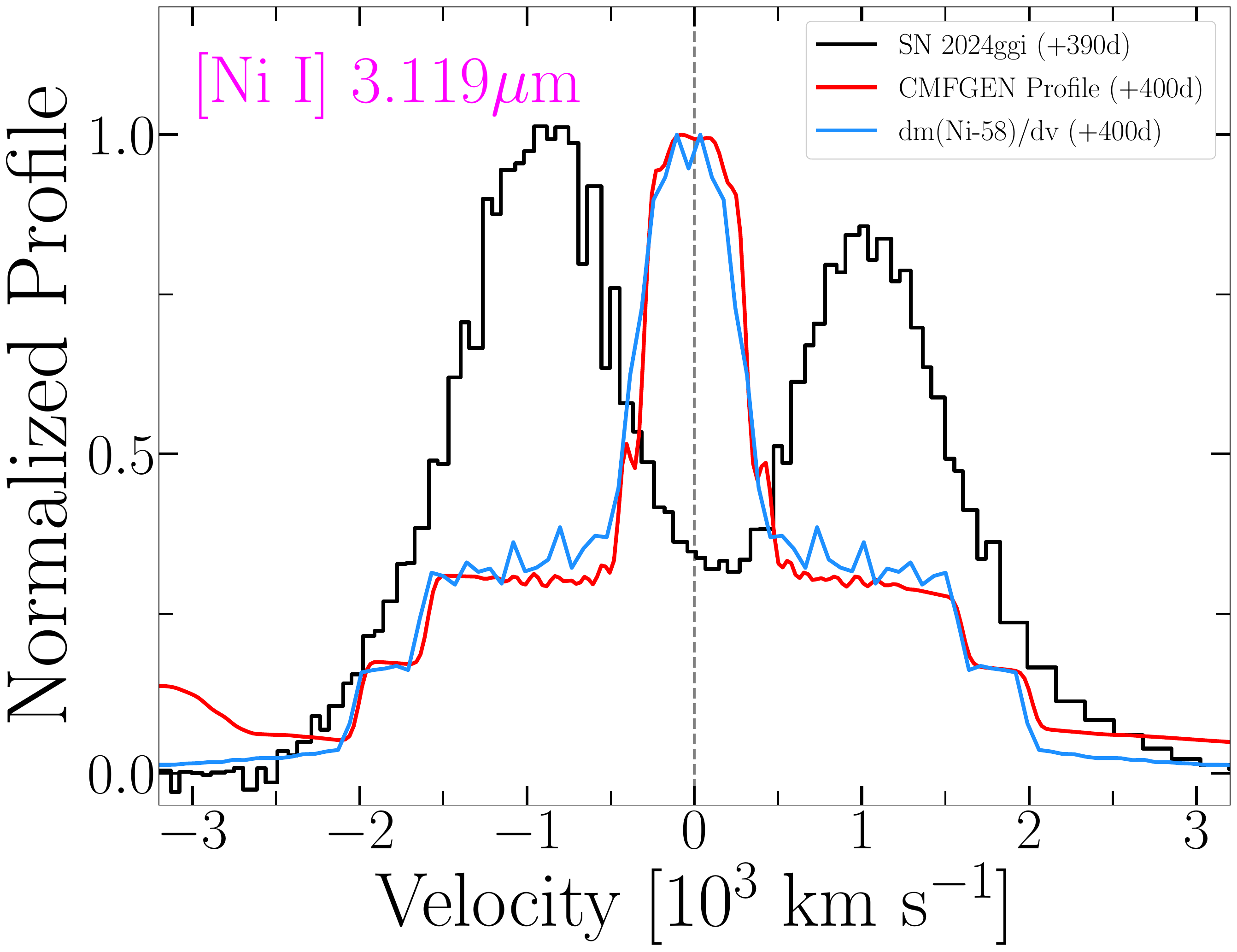}}
\subfigure{\includegraphics[width=0.33\textwidth]{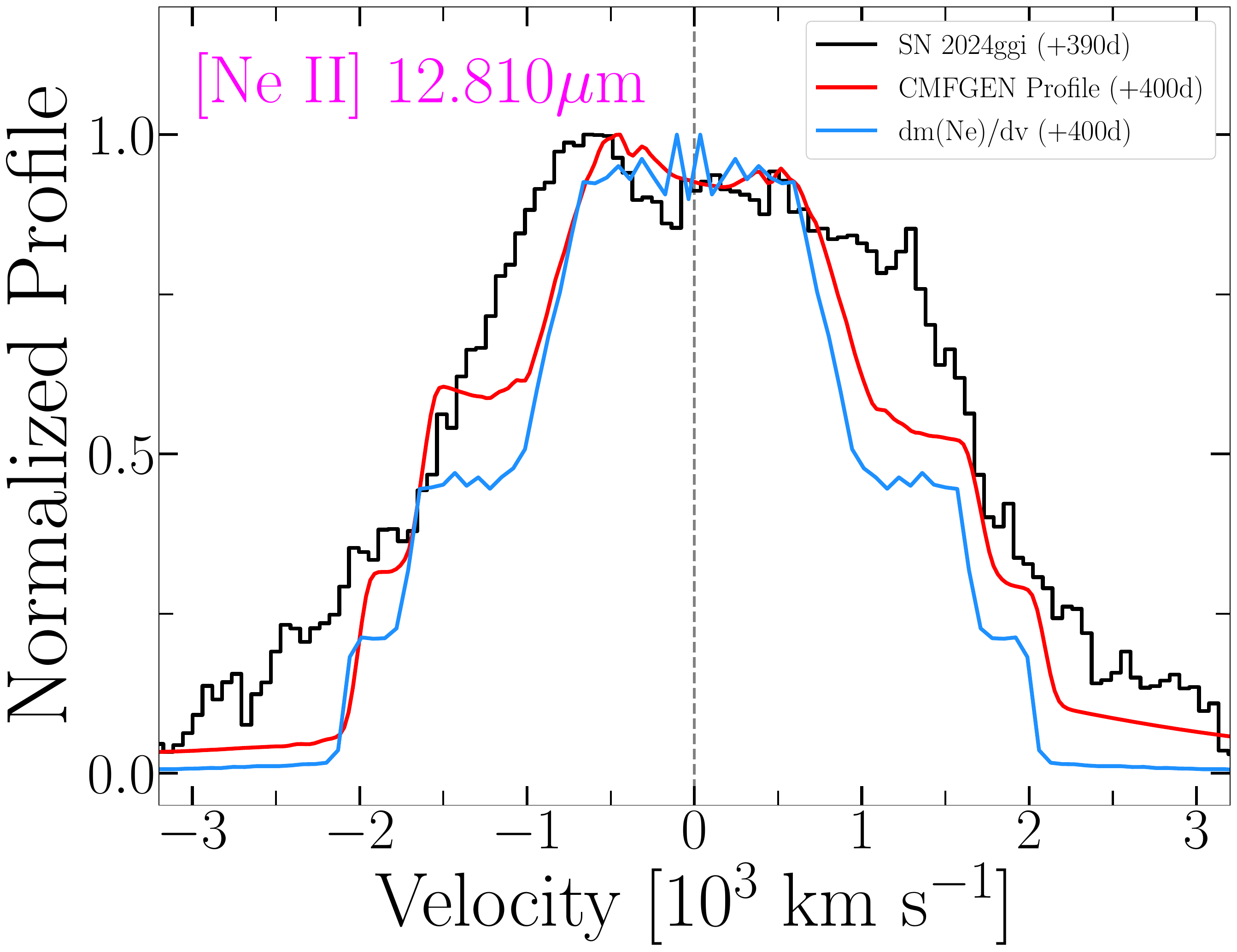}}
\subfigure{\includegraphics[width=0.33\textwidth]{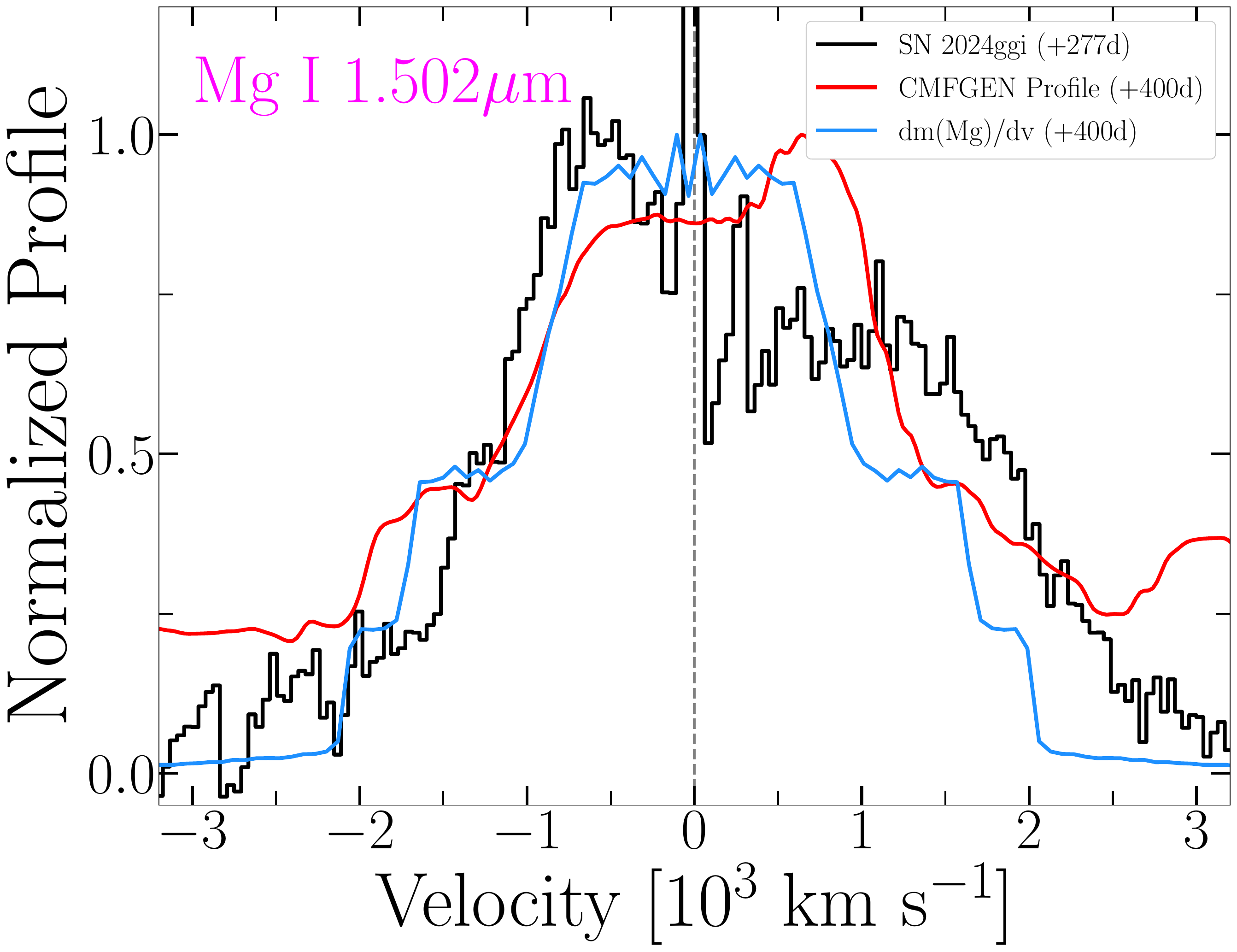}}
\caption{Emission line provide versus Doppler velocities of [\ion{Ni}{i}] $\lambda$3.119~$\mu$m ({\it left}), [\ion{Ne}{ii}] $\lambda$12.810~$\mu$m ({\it middle}), and [\ion{Mg}{i}] $\lambda$1.502~$\mu$m ({\it right}), all normalized to peak flux. SN~2024ggi shown in black compared to model predictions from {\tt CMFGEN} radiative transfer calculation (red) and the summing of ejecta mass per unit projected velocity ($dm/dv$; blue). Intriguingly, the latter can reproduce almost exactly the emission line profile produced through a non-LTE calculation in {\tt CMFGEN}.     \label{fig:cmfgen_v_dmdv} }
\end{figure*}

\subsection{Constructing Line Profiles without Non-LTE Calculations}\label{subsec:cmfgen}

Using the s15p2 model of \cite{Dessart21, Dessart25IR}, we conduct a simple experiment wherein we use the spatial distribution of various elements in the shuffled-shell ejecta structure to predict line profiles. The key, and perhaps controversial ansatz, is to assume that there is a direct mapping between element distribution and the corresponding element line emission. Here, we take the spherical s15p2 model and scan through all radial shells and polar angles, and sum the amount of mass at the corresponding projected velocity -- the SN ejecta is spherical so all directions are equivalent, and we thus take a viewing angle along the polar axis. Because the model is spherical, the distribution of mass per bin in projected velocity is symmetric. As shown in Figure \ref{fig:cmfgen_v_dmdv}, we repeat this process for various species including Ne, Mg, and Ni and show the resulting element distributions in projected velocity space together with the [\ion{Ni}{i}] $\lambda$3.1199~$\mu$m, [\ion{Ne}{ii}] $\lambda$12.8101~$\mu$m, and [\ion{Mg}{i}] $\lambda$1.50202~$\mu$m emission line profiles computed from a detailed radiative transfer calculation with \cmfgen. Strikingly, the two distributions agree closely and suggest that a similar calculation applied to 3-D explosion simulations would also provide a compelling profile morphology directly comparable to observations (e.g., see \S\ref{subsec:3dmodels}).

We, however, note that this exercise remains simplistic. For example, all emission lines from Ni or Mg should have exactly the same profile, whereas in practice, there is a fair amount of diversity, probably related to a distribution in density, ionization, or proximity to $^{56}$Ni blobs that are known to have complex distributions when examined in 3-D \citep{Vartanyan25, Giudici25}. Nonetheless, for the purpose of analyzing the maximum velocity attained by $^{56}$Ni created during explosive nucleosynthesis, this approach is robust. In a 3-D simulation at nebular times, the procedure would be similar to the calculation done with a 1-D, spherical output from \cmfgen. That is, all ejecta zones in {\it x,y,z} (the {\tt FLASH} hydrodynamics code uses a cartesian grid) are scanned and, for each zone, the calculation records the mass of a given element and its projected velocity. The ejecta being 3-D, the distribution will depend on viewing angle. If the cell {\it (x,y,z)} makes an angle $\theta$ to the distant observer, its projected velocity is just $V_r \cos\theta$. Running over all zones, one proceeds and sums all mass contributions falling into a predefined array of projected velocities. Summing over this distribution, one recovers the total yield of the corresponding element, as in the 1-D case applied to \cmfgen\ inputs.

\subsection{3-D Neutrino-Driven Explosion Model Comparisons}\label{subsec:3dmodels}

\begin{figure*}
\centering
\includegraphics[width=\textwidth]{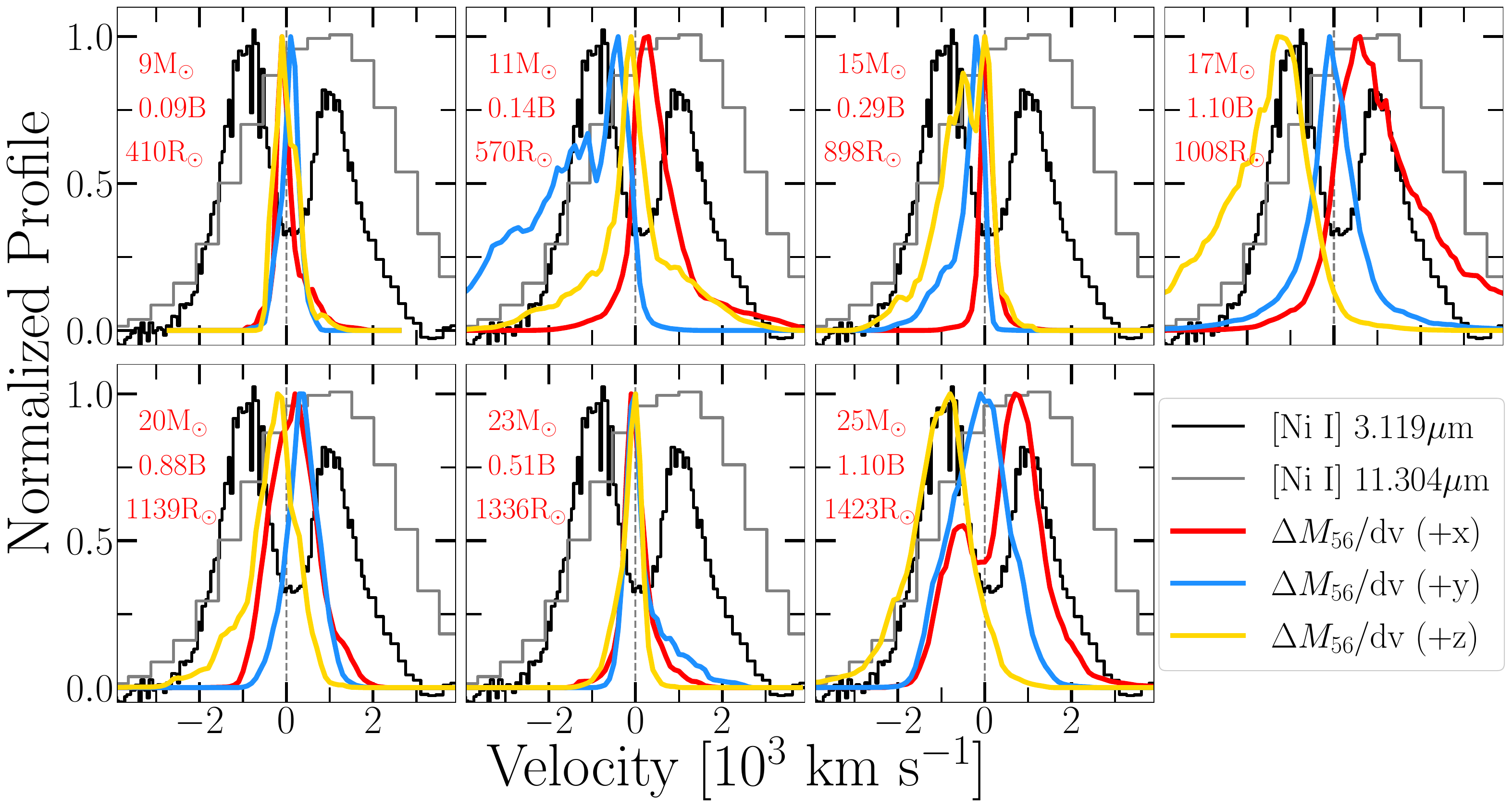}
\caption{ SN~2024ggi emission line velocities of [\ion{Ni}{ii}] $\lambda$11.304~$\mu$m (gray) and [\ion{Ni}{i}] $\lambda$3.119~$\mu$m (black) compared to the projected $^{56}$Ni distributions from 3-D neutrino-driven explosion models of $9-25~\Msun$ RSGs from \cite{Vartanyan25} computed along $+x$ (red), $+y$ (blue), and $+z$ (yellow) directions. Kinetic energy, initial radius and ZAMS mass per model shown in red. Only some of the viewing angles from the 11, 17, and 25~$\Msun$ models predict Ni velocities large enough to match the Ni distribution in SN~2024ggi and no model can match the symmetric, double-peaked structure of the observed [\ion{Ni}{i}] $\lambda$3.119~$\mu$m line. \label{fig:3D} }
\end{figure*}

3-D neutrino-driven explosion simulations now track the distribution of synthesized $^{56}$Ni from shock breakout until homologous expansion at $\sim$1-2~weeks after first light (e.g., \citealt{gabler_3dsn_21, Vartanyan25, Giudici25}). Based on the findings in \S\ref{subsec:cmfgen} for 1-D \cmfgen\ models, we conduct a similar experiment with the 3-D $^{56}$Ni distributions produced in neutrino-driven explosion simulations of $9-25~\Msun$ red supergiant progenitor star models from \cite{Vartanyan25}. In Figure \ref{fig:3D}, we present projected $^{56}$Ni velocities along $x$, $y$, and $z$ directions of the simulations, generated by summing $^{56}$Ni mass bins ($\Delta M_{56}$) per velocity bin ($dv$) at phases of $\sim 4 - 16$~days after shock breakout when the ejecta is in approximate homologous expansion. We compare these $^{56}$Ni velocities to the [\ion{Ni}{ii}] $\lambda11.304$~$\mu$m and [\ion{Ni}{i}] $\lambda3.119$~$\mu$m line profiles observed in SN~2024ggi for each progenitor model ZAMS mass. 

\begin{figure*}
\centering
\subfigure{\includegraphics[width=0.44\textwidth]{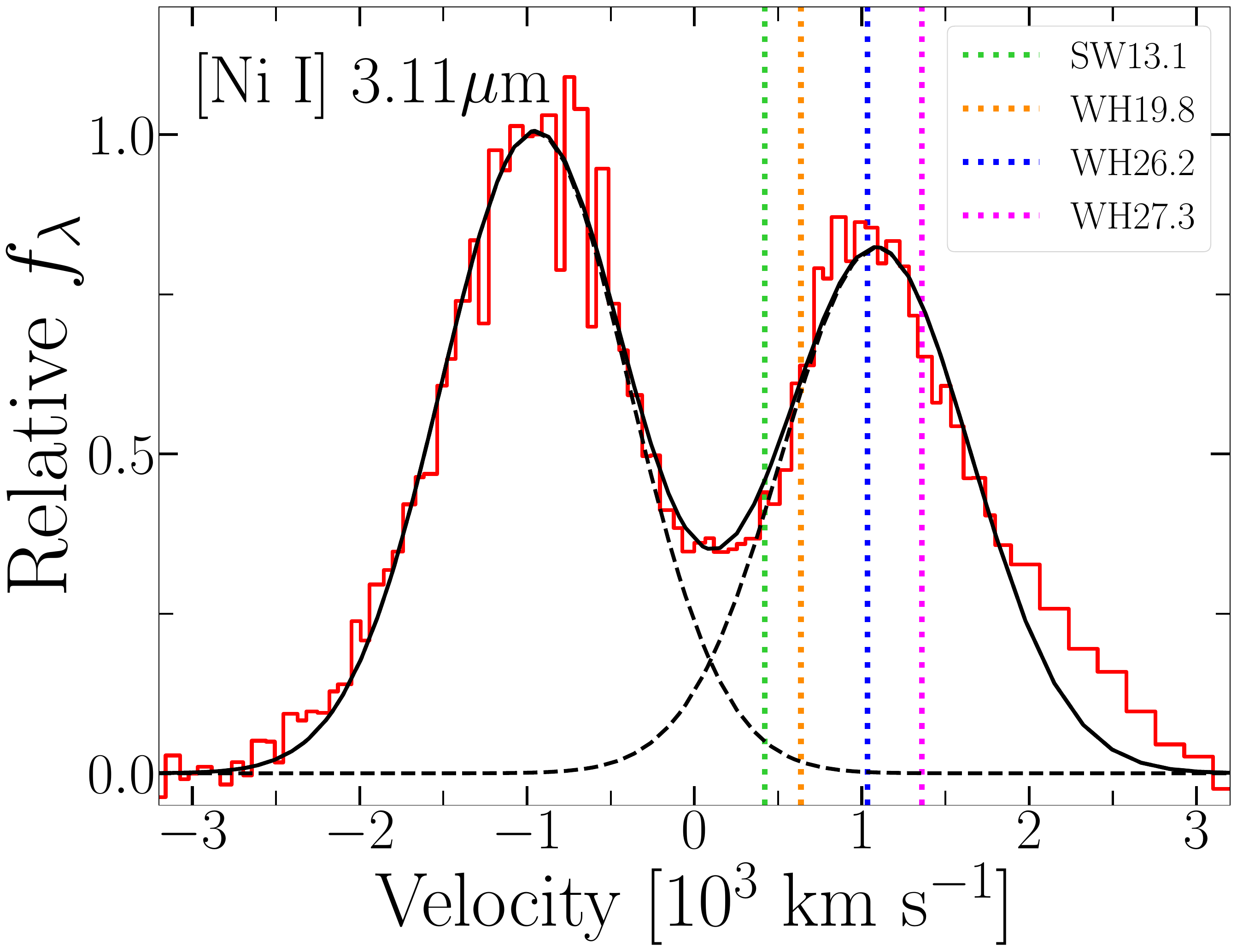}}
\subfigure{\includegraphics[width=0.55\textwidth]{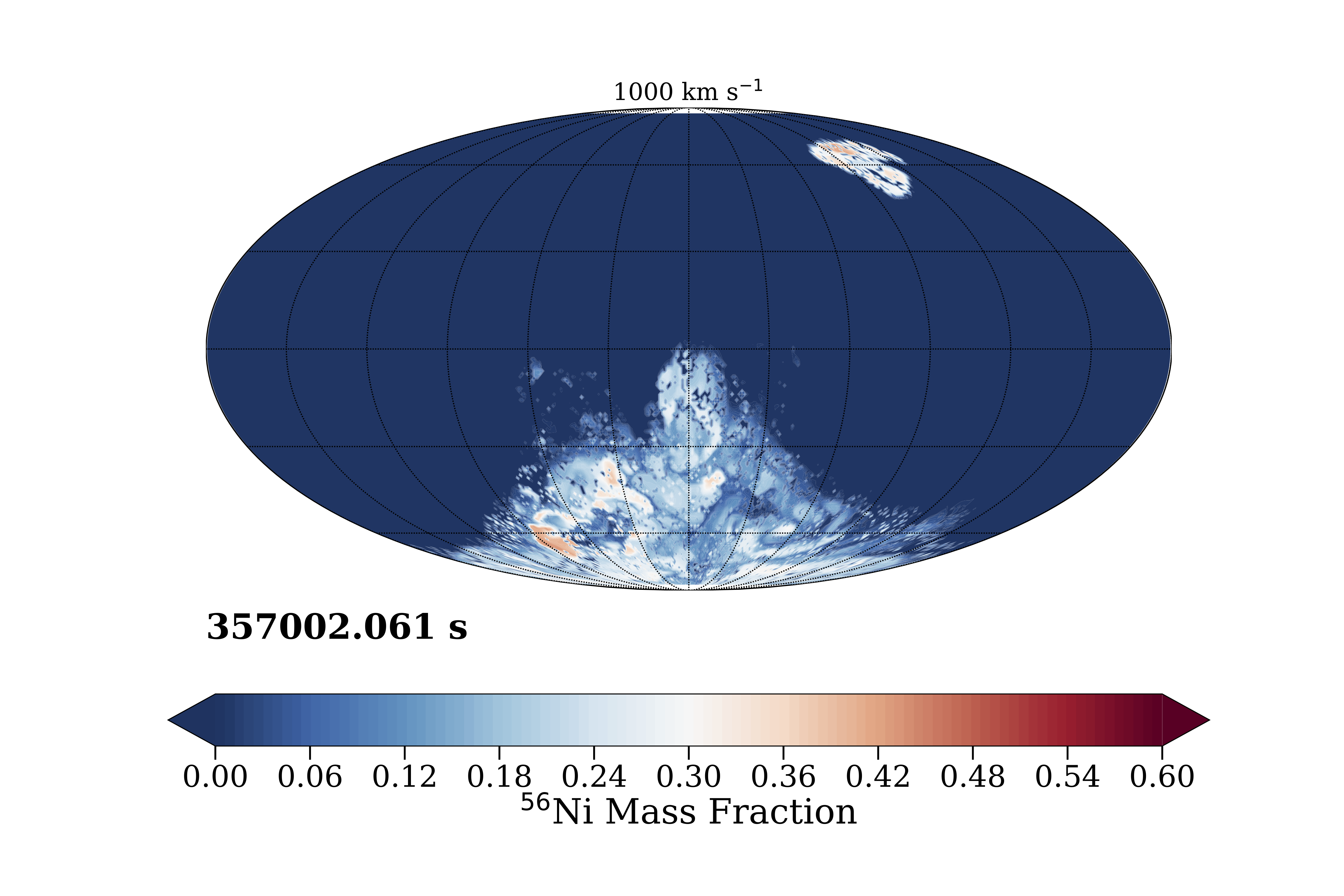}}\\
\subfigure{\includegraphics[width=0.44\textwidth]{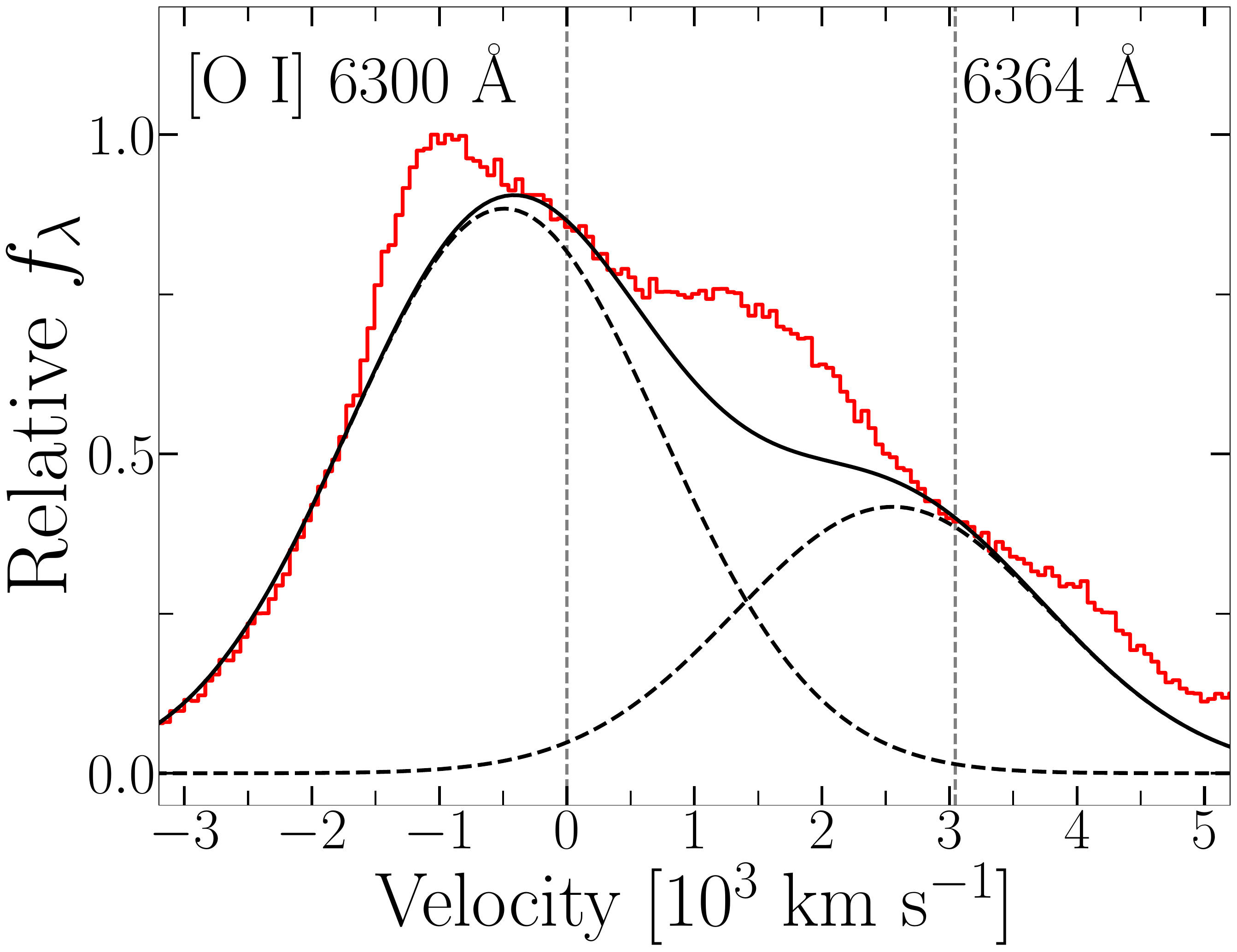}}
\subfigure{\includegraphics[width=0.55\textwidth]{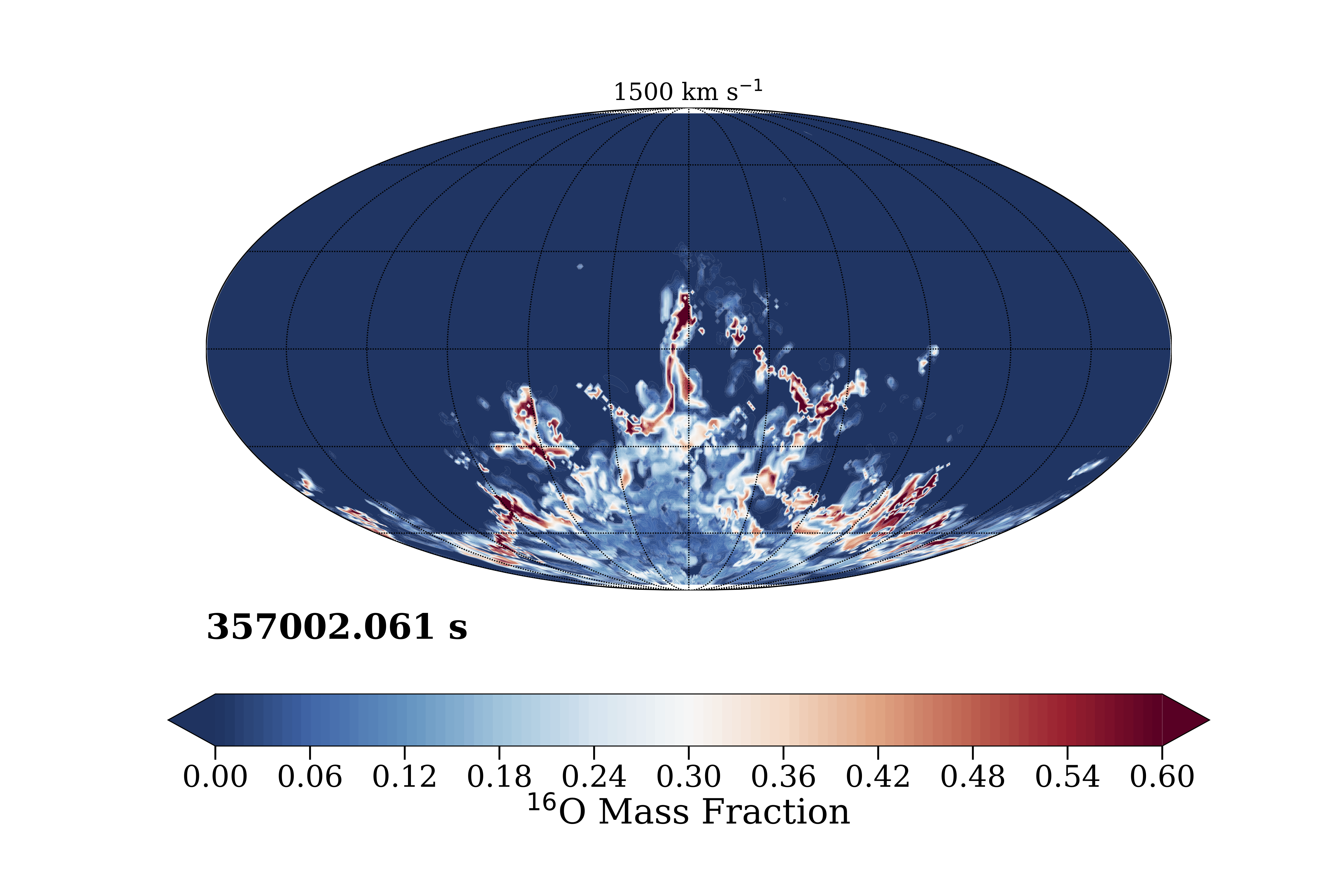}}\\
\caption{ {\it Left:} Observed [\ion{Ni}{i}] $\lambda$3.119~$\mu$m and [\ion{O}{i}] $\lambda$0.630~$\mu$m emission line velocities in SN~2024ggi (red), modeled with multiple Gaussian profiles (black dashed lines). The [\ion{O}{i}] doublet cannot be modeled with two Gaussians of the same FWHM, suggesting a more asymmetric ejecta distribution. Shown in dotted lines are the velocities corresponding to the peak of Ni/Co/Fe distributions at 10~days post-shock breakout found in the simulations by \cite{Giudici25} for 3-D explosions of $13.1 - 27.3~\Msun$ progenitor models. {\it Right:} Mollweide projection plots for 17~$\Msun$ 3-D explosion model \citep{Vartanyan25} created at ejecta velocity slices of $1000\,\kms$ for $^{56}$Ni (top) and $1500\,\kms$ for $^{16}$O (bottom) at 4.1~days after shock breakout.   \label{fig:moll} }
\end{figure*}

Overall, we find that the 11, 17 and 25~$\Msun$ 3-D explosion models of \cite{Vartanyan25} have large enough $^{56}$Ni velocities in at least one radial direction to match the velocities observed in the SN~2024ggi Ni line profiles. However, we note that only the 17 and 25~$\Msun$ models have the bulk of the $^{56}$Ni distribution at $\sim 1000~\kms$, consistent with SN~2024ggi. Notably, these models have the largest kinetic energies ($E_k =1.1$~B) and are comparable to the s15p2 model ($E_k = 0.84$~B), which also reproduces the extent of the observed Ni line velocities (e.g., Fig. \ref{fig:all}). Furthermore, only the 25~$\Msun$ model contains a double-peaked structure (when viewed in the $+x$ direction) similar to the [\ion{Ni}{i}] $\lambda3.119$~$\mu$m emission line profile. We also compare the spread in Ni velocities observed in SN~2024ggi to 3-D neutrino-driven explosion models from \cite{Giudici25} in the upper left panel of Figure \ref{fig:moll}. Here, we plot the radial velocities corresponding to the peak of the Ni/Co/Fe mass distribution at 10~days post-explosion for different progenitor model masses (see their Fig. 11) with respect to the [\ion{Ni}{i}] $\lambda3.119$~$\mu$m profile. Notably, the bulk of the Ni/Co/Fe mass matches the peak of the [\ion{Ni}{i}] profile ($\sim1000~\kms$) for models $>20~\Msun$ e.g., the $26.2~\Msun$ model matches the peak exactly. However, $\Delta M_{56}/dv$ was not explicitly calculated for these models so comparison similar to that shown for the models of \cite{Vartanyan25} cannot be performed at this time. 

In Figure \ref{fig:moll}, we further visualize the asymmetries present in the 3-D models of \cite{Vartanyan25} through Mollweide projection plots for the distribution of $^{56}$Ni at $1000~\kms$ and also $^{16}$O at $1500~\kms$ in the 17~$\Msun$ explosion model.\footnote{Mollweide plots for all 3-D models can be found here: \url{https://github.com/wynnjacobson-galan/Type-II-IR-Lines} } Based on the double-peaked profiles observed in some IGE transitions, SN~2024ggi likely has a bipolar and/or asymmetric distribution of $^{56}$Ni similar to the 3-D explosion models, but with the main difference being the mass ratios of each ``lobe'' of material. As seen in the 17~$\Msun$ model, the $^{56}$Ni is distributed in a bipolar shape, but with one lobe being more massive than the other, leading to asymmetric line profiles that do not match the [\ion{Ni}{i}] $\lambda3.119$~$\mu$m profile in SN~2024ggi. A similar distribution of $^{16}$O at $1500~\kms$ can be seen in the Mollweide plot, which we also compare to the asymmetric structure observed in [\ion{O}{i}] $\lambda0.630$~$\mu$m.

\section{Discussion} \label{sec:discussion}

Examination of nebular emission lines from optical to mid-IR wavelengths enables us to reconstruct the inner ejecta geometry of SN~2024ggi and probe the products of explosive nucleosynthesis. Overall, one of the most intriguing observables in SN~2024ggi is the double-peaked and/or asymmetric profiles that are present in some IGE and IME emission lines. As shown in Figures \ref{fig:IGEvels} and \ref{fig:2D}, significant double-peaked structure is observed in prominent IGE emission lines such as [\ion{Ni}{i}] $\lambda$3.119~$\mu$m, [\ion{Co}{i}] $\lambda$12.255~$\mu$m and [\ion{Fe}{ii}] $\lambda$17.931~$\mu$m. However, the ratio of blue vs red peaks in these IGE features is close to unity, indicating a similar distribution of material in projected velocity space. Notably, some Ni and Co lines do not show an obvious double-peaked structure (e.g., [\ion{Ni}{i}] $\lambda$11.304~$\mu$m, [\ion{Ni}{ii}] $\lambda$1.939~$\mu$m, [\ion{Co}{ii}] $\lambda$10.520~$\mu$m), but the low resolution of key Ni lines such as [\ion{Ni}{ii}] $\lambda$6.634~$\mu$m and [\ion{Ni}{i}] $\lambda$7.505~$\mu$m prohibits a complete census of Ni line morphology. 

Specifically, the different shapes of \ion{Ni}{i} lines (e.g., $\lambda$3.119~$\mu$m vs. $\lambda$11.304~$\mu$m) can be explained by the temperature and ionization stratification across the inner ejecta. Here, [\ion{Ni}{i}] $\lambda$3.119~$\mu$m likely traces a bi-polar inner core region containing shielded, high density regions of Ni given that it requires higher temperatures for excitation ($\Delta E = 0.397$~eV). Conversely, the Gaussian-shaped [\ion{Ni}{i}] $\lambda$11.304~$\mu$m, centered red-ward at $\sim1000~\kms$, traces lower density gas at larger radii and has a lower temperature sensitivity ($\Delta E = 0.11$~eV). [\ion{Ni}{ii}] $\lambda$1.939~$\mu$m shows a similar Gaussian morphology, suggesting an ionized emitting region that is centrally filled than the [\ion{Ni}{i}] $\lambda$3.119~$\mu$m emission region. Beyond the bipolar central IGE region, IMEs such as Ne, O, Ar and Mg also show evidence for ejecta asymmetries at higher velocities, although their double-peaked structure is not as defined as the IGE emission lines. Overall, modeling of the double-peaked profiles with varying ejecta distributions (see \S\ref{subsubsection:density}) suggests that some [\ion{Ni}{i}] emission arises from regions with the most extreme polar enhancement ($a_1 = 7$), with [\ion{Co}{i}] and [\ion{Fe}{ii}] also arises from regions where polar densities are increased by factors of $\sim 2-4$. 

The differing line morphologies among the iron-peak elements indicate that neutral and ionized components do not trace identical emitting volumes within the inner ejecta. The double-peaked profiles observed in [\ion{Ni}{i}] $\lambda$3.119~$\mu$m and $\lambda$11.998~$\mu$m, together with the similarly double-peaked [\ion{Fe}{ii}] $\lambda$1.644~$\mu$m and $\lambda$17.931~$\mu$m lines and the [\ion{Co}{i}] $\lambda$12.255~$\mu$m transition, point to a bipolar distribution of cooler, recombined material. In contrast, the singly ionized species — most notably [\ion{Ni}{ii}] $\lambda$1.939~$\mu$m and [\ion{Co}{ii}] $\lambda$10.520~$\mu$m — as well as the [\ion{Ni}{i}] $\lambda$7.505~$\mu$m line, exhibit broader, more centrally filled and red-shifted profiles. This suggests that these transitions arise from a physically distinct component that is either more extended and/or weighted toward the receding side of the ejecta. 

Taken together, these observations imply that the iron-rich ejecta core contains stratification in geometry, temperature and ionization. The neutral species (e.g., \ion{Ni}{i}, \ion{Co}{i}) and low-excitation \ion{Fe}{ii} lines preferentially trace cooler, denser bipolar lobes, while the singly ionized species trace a warmer and more ionized medium that is more volume-filling and less strongly bi-modal in velocity space. The similarity between the profiles of [\ion{Ni}{ii}] $\lambda$1.939~$\mu$m, [\ion{Co}{ii}] $\lambda$10.520~$\mu$m, and [\ion{Ni}{i}] $\lambda$7.505~$\mu$m further suggests that ionization and excitation effects, rather than elemental abundance alone, play a dominant role in shaping the observed line profiles. Overall, the observed differences in line profile morphology across ionization states is consistent with an inner ejecta geometry shaped by the $^{56}$Ni bubble effect. Here, neutral double-peaked species such as [\ion{Ni}{i}] and [\ion{Co}{i}] would trace the denser, lower temperature edges of the bubbles while emission from the ionized states e.g., [\ion{Ni}{ii}], [\ion{Co}{ii}] would arise from from higher temperature regions within the bubbles. Furthermore, emission from IMEs such as Mg, Ne and O would form within the compression regions between bubbles, giving rise to their asymmetric line morphologies, in addition to forming at larger velocities.  

\section{Conclusions} \label{sec:conclusion}

In this paper, we present analysis and modeling of the nebular optical/IR spectra of SN~II 2024ggi. Below we summarize the primary observational and theoretical findings from this work.  

\begin{itemize}

    \item The nebular optical-IR spectra of SN~2024ggi contain emission lines of iron-group and intermediate-mass elements with diverse morphologies. Direct comparison and cross-correlation analysis of Ni, Co and Fe shows that the double-peaked lines (e.g., in [\ion{Ni}{i}], [\ion{Co}{i}], and [\ion{Fe}{ii}]) are closely aligned in velocity space, with peak separations and centroids consistent to within $\sim 100–200~\kms$, indicating a shared large-scale structure. In contrast, lines with Gaussian or skewed profiles ([\ion{Ni}{ii}]~$\lambda$1.939~$\mu$m, [\ion{Co}{ii}]~$\lambda$10.520~$\mu$m, [\ion{Ni}{i}]~$\lambda$11.304~$\mu$m) exhibit systematic velocity offsets, with best-fit lags of order several hundred $\kms$, implying emission from physically distinct regions. Resolved IME lines of Ne, Mg, O and Ca also show asymmetric structure and in some cases weakly double-peaked profiles, but with less pronounced peak separation and asymmetry than the IGEs, suggesting a more extended and less strongly structured distribution. 

    \item The combination of line profiles provides strong evidence for a bipolar geometry and/or clumping of the inner metal-rich ejecta, as indicated by the consistent double-peaked structure observed in neutral species such as [\ion{Ni}{i}] $\lambda$3.119~$\mu$m and [\ion{Co}{i}] $\lambda$12.255~$\mu$m as well as low-excitation Fe lines (e.g., [\ion{Fe}{ii}] $\lambda$1.644~$\mu$m and $\lambda$17.931~$\mu$m). In contrast, singly-ionized transitions (e.g., [\ion{Ni}{ii}]~$\lambda$1.939~$\mu$m, [\ion{Co}{ii}]~$\lambda$10.520~$\mu$m) and select neutral lines (e.g., [\ion{Ni}{i}]~$\lambda$7.505~$\mu$m and $\lambda$11.304~$\mu$m) exhibit smoother, centrally filled or red-skewed profiles, indicating that these species preferentially trace a more volume-filling, higher-ionization component. This dichotomy demonstrates that the underlying bipolar Ni distribution is modulated by stratification of ionization, temperature and density, all of which are consistent with expectations for the $^{56}$Ni bubble effect.

    \item LTE-based estimates yield total Ni and Co masses of $M_{\rm Ni} \approx 2.1 \times 10^{-3}~\Msun$ and $M_{\rm Co} \approx 1.4 \times 10^{-3}~\Msun$, but electron densities comparable to critical densities imply that these values should be treated as lower limits due to non-LTE effects. Using a $^{56}$Ni mass of $M(^{56}{\rm Ni}) = 0.05~\Msun$ as inferred from light curve modeling results in a radioactive to stable Ni mass ratio of $\lesssim 38$.

    \item Comparison of near- and mid-IR IGE (e.g., Ni) and IME (e.g., Ne, Mg, Ar) emission line luminosities to \cmfgen\ model spectra favors a progenitor model with $M_{\rm ZAMS} \approx 12–15~\Msun$, consistent with constraints using the [\ion{O}{i}]~$\lambda$0.630~$\mu$m optical emission line strength as well as light curve modeling and pre-explosion analysis. Overall, the s15p2 model ($M_{\rm ZAMS}=15.2~\Msun$, $E_k = 0.84$~B) from \citep{Dessart25IR} accurately reproduces the line profiles of the IR H and Ne lines in SN~2024ggi as well as the velocity spread observed in IGE profiles.

    \item Using a 1-D explosion model of a $15.2~\Msun$ progenitor, we calculate line profiles of Ni, Ne and Mg through both a detailed non-LTE radiative transfer calculation with \cmfgen\ \citep{Dessart25IR} and by summing radial shells of ejecta ($\Delta M$) per unit of projected velocity ($dv$). We find consistent line morphology across methods, suggesting a direct mapping between ejecta mass and line emissivity, without the explicit need for a full non-LTE radiative transfer calculations. This method provides a representative observable from core-collapse simulations that can be confronted with optical/IR nebular spectroscopy of SNe~II.

    \item We calculate projected velocity profiles of $^{56}$Ni along three lines of sight ($+x,+y,+z$) using the 3-D neutrino-driven explosion simulations of \cite{Vartanyan25}. From this $9-25~\Msun$ model grid, we find that only the 11, 17 and 25~$\Msun$ model predicts Ni-mixing out to large enough velocities to match the Ni line profiles observed in SN~2024ggi. However, the distribution of Ni is too asymmetric in these simulations compared to the bipolar ejecta structure inferred from the observed double-peaked structure in Ni, Co, and Fe emission lines. 
    
\end{itemize}

Nebular IR spectroscopy of SNe~II such as SN~2024ggi with {\it JWST} provides a direct probe of the 3-D structure of the inner ejecta. The detection of isolated Ni, Co and Fe emission lines in the IR is essential to robustly mapping explosive nucleosynthesis and constraining the composition of the inner, Ni-rich regions. To date, only two SNe~II (2023ixf and 2024ggi) have nebular mid-IR spectra with {\it JWST}, but neither SN has completely resolved mid-IR emission lines of all IGEs. Consequently, it is imperative that nebular, medium-resolution spectroscopy of more SNe~II be obtained with {\it JWST} in order to constrain the diversity of line profile morphology and for robust modeling with both 3-D neutrino-driven explosion simulations and non-LTE radiative transfer codes. 

\begin{deluxetable*}{ccccccccccc}[t!] 
\tablecaption{SN~2024ggi Emission Line Properties \label{tab:lines}}
\tablehead{
\colhead{Ion} & \colhead{$\lambda$} & \colhead{$\Delta E$} & \colhead{$A_{ul}$} & \colhead{Temp. Sensitivity} & \colhead{$n_{\rm crit}^a$} & \colhead{Profile Shape} & \colhead{Line Lum.} & \colhead{Phase} & \colhead{$f_B^b$} & \colhead{$f_R^c$}\\
\colhead{} & \colhead{($\mu$m)} & \colhead{(eV)} & \colhead{(s$^{-1}$)} & \colhead{} & \colhead{(cm$^{-3}$)} & \colhead{} & \colhead{($10^{37}$~erg~s$^{-1}$)} & \colhead{(days)} & \colhead{} & \colhead{} }
\startdata
Ni I  & 3.120  & 0.397 & $7.8\times10^{-2}$  & Medium & $4.5\times10^{6}$ & Double-peaked & $2.23 \pm 0.12$ & 286 & $0.51 \pm 0.01$ & $0.49 \pm 0.01$ \\
Ni I  & 3.120  & 0.397 & $7.8\times10^{-2}$  & Medium & $4.5\times10^{6}$ & Double-peaked & $3.19 \pm 0.18$ & 387 & $0.53 \pm 0.01$ & $0.47 \pm 0.01$ \\
Ni I  & 7.505 & 0.165 & $6.2\times 10^{-2}$  & Low & $2.6 \times 10^{6}$ & Gaussian & $10.2 \pm 0.56$ & 265 & -- & -- \\
Ni I  & 7.505 & 0.165 & $6.2\times 10^{-2}$  & Low & $2.6 \times 10^{6}$ & Gaussian & $3.82 \pm 0.21$ & 390 & -- & -- \\
Ni I  & 11.307 & 0.110 & $2.5\times10^{-2}$  & Low  & $1.4\times10^{6}$ & Gaussian, red-skewed & $2.84 \pm 0.16$ & 265 & -- & -- \\
Ni I  & 11.307 & 0.110 & $2.5\times10^{-2}$  & Low  & $1.4\times10^{6}$ & Gaussian, red-skewed & $2.08 \pm 0.11$ & 390 & -- & -- \\
Ni I  & 12.001 & 0.103 & $2.1\times10^{-2}$  & Low  & $8.6\times10^{5}$ & Double-peaked? & $0.45 \pm 0.025$ & 390 & -- & -- \\
Ni II & 1.939  & 0.639 & $8.7\times10^{-2}$  & High & $7.1\times10^{6}$ & Gaussian, red-skewed? & $8.23 \pm 0.47$ & 286 & -- & -- \\
Ni II & 1.939  & 0.639 & $8.7\times10^{-2}$  & High & $7.1\times10^{6}$ & Gaussian, red-skewed? & $3.07 \pm 0.17$ & 387 & -- & -- \\
Ni II & 6.636  & 0.187 & $5.5\times10^{-2}$ & Low  & $2.7\times10^{6}$ & Gaussian & $19.9 \pm 1.10$ & 266 & -- & -- \\
Ni II & 6.636  & 0.187 & $5.5\times10^{-2}$ & Low  & $2.7\times10^{6}$ & Gaussian & $23.3 \pm 1.29$ & 390 &  -- & -- \\
Co I  & 12.26 & 0.101 & $1.9\times 10^{-2}$ & Low & $\sim 1.3 \times 10^5$ & Double-peaked & $1.31 \pm 0.10$ & 390 & $0.49 \pm 0.02$ & $0.51 \pm 0.02$ \\
Co II  & 10.52 & 0.118 & $2.2 \times 10^{-2}$  & Low  & $\sim 1.3 \times 10^5$ & Gaussian & $5.60 \pm 0.31$ & 266 &  -- & -- \\
Co II  & 10.52 & 0.118 & $2.2 \times 10^{-2}$  & Low  & $\sim 1.3 \times 10^5$ & Gaussian & $2.82 \pm 0.16$ & 390 &  -- & -- \\
Fe II  & 1.644 & 0.754 & $4.6\times10^{-3}$ & High & $3.0\times10^{4}$ & Double-peaked & $18.9 \pm 1.0$ & 277 & $0.58 \pm 0.04$ & $0.42 \pm 0.04$ \\
Fe II  & 1.644 & 0.754 & $4.6\times10^{-3}$ & High & $3.0\times10^{4}$ & Double-peaked & $20.7 \pm 1.1$ & 400 & $0.40 \pm 0.02$ & $0.60 \pm 0.02$ \\
Fe II & 17.931 & 0.069 & $6.0\times10^{-3}$ & Low & $3.9\times10^{4}$ & Double-peaked & $2.97 \pm 0.16$ & 266 & $0.53 \pm 0.08$ & $0.47 \pm 0.08$ \\
Ar II & 6.983 & 0.178 & $5.3\times10^{-2}$ & Low & $8.7\times10^{5}$ & Gaussian & $4.04 \pm 0.22$ & 266 &  -- & -- \\
Ar II & 6.983 & 0.178 & $5.3\times10^{-2}$ & Low & $8.7\times10^{5}$ & Gaussian & $2.98 \pm 0.16$ & 390 &  -- & -- \\
Mg I & 1.502 & 0.825 & $2.4\times10^{-1}$ & High & $5.9\times10^{6}$ & Double-peaked & $22.6 \pm 1.25$ & 277 & $0.31 \pm 0.05$ & $0.69 \pm 0.05$ \\
Mg I & 1.502 & 0.825 & $2.4\times10^{-1}$ & High & $5.9\times10^{6}$ & Double-peaked? & $7.2 \pm 0.40$ & 400 & -- & -- \\
Mg I & 3.867 & 0.321 & $8.6\times10^{-2}$ & Medium & $2.1\times10^{6}$ & Double-peaked & $2.26 \pm 0.13$ & 286 & $0.26 \pm 0.01$ & $0.74 \pm 0.01$ \\
Mg I & 3.867 & 0.321 & $8.6\times10^{-2}$ & Medium & $2.1\times10^{6}$ & Double-peaked & $0.70 \pm 0.04$ & 387 & $0.30 \pm 0.02$ & $0.70 \pm 0.02$ \\
Ne II & 12.81 & 0.097 & $8.6\times10^{-3}$ & Low & $1.4\times10^{5}$ & Double-peaked & $5.04 \pm 0.28$ & 266 & $0.46 \pm 0.06$ & $0.54 \pm 0.06$ \\
Ne II & 12.81 & 0.097 & $8.6\times10^{-3}$ & Low & $1.4\times10^{5}$ & Double-peaked & $4.53 \pm 0.25$ & 390 & $0.55 \pm 0.06$ & $0.45 \pm 0.06$ \\
\enddata
\tablenotetext{a}{Critical densities estimated assuming electron collisions dominate and
$n_{\rm crit}=A_{ul}/q_{ul}$ with $q_{ul}=8.63\times10^{-6}\Omega/(g_u T^{1/2})$,
evaluated at $T_e=5000$ K and collision strength $\Omega=1$.}
\tablenotetext{b}{$f_B = \frac{L_B |v_{\rm B, peak}|}{L_1 |v_{\rm 1, peak}| + L_R |v_{\rm R, peak}|}$}
\tablenotetext{c}{$f_R = \frac{L_R |v_{\rm R, peak}|}{L_B |v_{\rm B, peak}| + L_R |v_{\rm R, peak}|}$}
\end{deluxetable*}

\facilities{\emph{James Webb Space Telescope}, Keck I/II (LRIS, NIRES)}

\software{\cmfgen\ \citep{hillier12, dessart22}}

\section{Acknowledgments} \label{Sec:ack}

We thank Stan Barmentloo for engaging discussions and assistence with Einstein coefficients. W.J.-G.\ is supported by NASA through Hubble Fellowship grant HSTHF2-51558.001-A awarded by the Space Telescope Science Institute, which is operated for NASA by the Association of Universities for Research in Astronomy, Inc., under contract NAS5-26555. This work was granted access to the HPC resources of TGCC under the allocation 2024 -- A0170410554 made by GENCI, France. Some/all of the data presented in this article were obtained from the Mikulski Archive for Space Telescopes (MAST) at the Space Telescope Science Institute. The specific observations analyzed can be accessed via \dataset[doi:10.17909/6wfn-2f02]{https://doi.org/10.17909/6wfn-2f02}.

\bibliographystyle{aasjournal} 
\bibliography{references} 


\end{document}